\documentclass[aps,prl,reprint,superscriptaddress,nofootinbib,floatfix,preprintnumbers]{revtex4-2}

\usepackage{amsmath,amssymb,bm,mathtools}
\usepackage{graphicx}
\usepackage{xcolor}
\usepackage[colorlinks=true,allcolors=blue]{hyperref}
\usepackage{microtype}
\usepackage{placeins}
\usepackage{tikz}
\usepackage{booktabs}
\usepackage{float}
\usetikzlibrary{arrows.meta,calc,decorations.pathreplacing}

\newcommand{\tr}{\operatorname{tr}}
\newcommand{\Tr}{\operatorname{Tr}}
\newcommand{\ii}{\mathrm{i}}

\newcommand{\bea}{\begin{eqnarray}}
\newcommand{\eea}{\end{eqnarray}}

\begin{document}

\preprint{YITP-26-122}

\title{Timelike Entanglement from Spacetime Density Matrices: A Lattice Realization}

\author{Hao-yu Wang}
\email{haoyuwong777@gmail.com}
\affiliation{School of Physics, Huazhong University of Science and Technology, Luoyu Road 1037, Wuhan, Hubei 430074, China}
\author{Wu-zhong Guo}
\email{wuzhong@hust.edu.cn}

\affiliation{School of Physics, Huazhong University of Science and Technology, Luoyu Road 1037, Wuhan, Hubei 430074, China}
\affiliation{Center for Gravitational Physics and Quantum Information,
Yukawa Institute for Theoretical Physics, Kyoto University, Kyoto 606-8502, Japan}

\date{\today}

\begin{abstract}
We develop a systematic framework for timelike entanglement in a
two-dimensional free real scalar field, connecting spacetime density
matrices with Lorentzian twist-operator correlators through independent
microscopic and continuum calculations. Extending Gaussian diagonalization
to generally non-Hermitian reduced operators, we determine their complete
nonzero spectra in the generic regular case, together with all integer
R\'enyi moments. The real-time replica construction relates these moments
to twist correlators, with quantitative agreement in both magnitude and
phase for the full four-point function on a circle, the Dirichlet and
Neumann boundary two-point functions, and the massive form-factor
predictions. In the massless continuum theory with Dirichlet boundaries,
we further establish a finite causally connected window in which all
integer moments are real, while the even moments are strictly positive for
arbitrary initial states. For Gaussian states, all integer moments are
strictly positive, and the resulting R\'enyi entropies are real.
These results demonstrate that causal connectedness alone is not
sufficient for a nonzero imaginary part of timelike R\'enyi entropy.
\end{abstract}

\maketitle

\section{Introduction}

Entanglement entropy, defined from the reduced density matrix of a
spatial subregion on a Cauchy surface, has become a standard probe in
quantum many-body physics, quantum field theory (QFT), and gravity
\cite{Amico:2007ag,Calabrese:2009qy,Zeng:2015pxf,Rangamani:2016dms}.
In QFT, the replica method provides a central analytic framework for
studying entanglement. In two-dimensional QFTs, it represents R\'enyi
entropies by partition functions on branched geometries
\cite{Calabrese:2004eu}, which can equivalently be expressed as
correlation functions of branch-point twist fields in the replicated
theory~\cite{Cardy:2007mb}; see~\cite{Doyon:2025xvo} for a recent
review. This representation has led to a deeper understanding of
entanglement in both QFT and quantum many-body systems. In parallel,
lattice methods provide a microscopic approach, allowing entanglement
to be computed directly from a regulated reduced density matrix and
compared with continuum replica predictions
\cite{Bombelli:1986rw,Srednicki:1993im,Callan:1994py,Vidal:2002rm,
peschel2003calculation,Casini:2009sr}.

While this conventional formulation concerns spatial subregions on a
single Cauchy surface, relativistic theories naturally motivate a
broader framework in which spatial and temporal correlations are treated
on an equal footing. Various such frameworks have been proposed
\cite{aharonov1964time,Aharonov:2007czt,Marcovitch:2011vfd,
Sorkin:2012sn,Fitzsimons:2013gga,Buscemi:2013xlk,Leifer:2013ghr,
Horsman:2017nqa,Cotler:2017anu,Chen:2020ild,Diaz:2020dfe,
Fullwood:2022rjd,Parzygnat:2022pax,Lie:2024kbl,Milekhin:2025ycm,
Guo:2025dtq,Das:2025fcd,Diaz:2026qrf,Jones:2026vkw}. Alongside these
developments, various notions of timelike or temporal entanglement have
also been explored
\cite{Brukner:2004egd,Fritz:2010qzm,Olson:2010jy,Hastings:2014qqa,
Wang:2018jva,Lerose:2021svg,Giudice:2021smd,Foligno:2023dih,
Liu:2022ugc,Doi:2022iyj,Carignano:2023xbz,Bou-Comas:2024pxf,
Milekhin:2025ycm,Guo:2025dtq}.


From the perspective of QFT, an operator that reproduces Wightman
correlation functions provides a well-defined notion of entanglement in
time and admits a natural Schwinger--Keldysh (SK) path-integral
representation~\cite{Milekhin:2025ycm}. Such a construction can be
formulated in terms of a spacetime density matrix associated with two or
more Cauchy surfaces~\cite{Guo:2025dtq}. Tracing out complementary
degrees of freedom then defines reduced spacetime density matrices and
the corresponding entropy-related quantities.

More precisely, let \(C_0\) and \(C_t\) be Cauchy surfaces at times \(0\)
and \(t\), respectively, with initial state \(\rho_0\) and time-evolution
operator \(U=e^{-\ii Ht}\). The full spacetime density
matrix \(T_{C_0C_t}\) is defined by
\begin{equation}
    \tr\!\left[
        T_{C_0C_t}(O_0\otimes O_1)
    \right]
    =
    \Tr\!\left[
        \rho_0 O_0 U^\dagger O_1 U
    \right],
    \label{eq:full-spacetime-density-matrix}
\end{equation}
for arbitrary $O_0$ and $O_1$. For subregions \(A_0\subset C_0\) and \(B_t\subset C_t\), tracing out the complementary degrees of freedom defines the reduced spacetime density
matrix
\begin{equation}
    T_{A_0B_t}
    =
    \tr_{\bar A_0\bar B_t}T_{C_0C_t},
    \label{eq:reduced-spacetime-density-matrix}
\end{equation}
where \(\bar A_0\) and \(\bar B_t\) denote the complements of \(A_0\)
and \(B_t\), respectively. In general, neither \(T_{C_0C_t}\) nor
\(T_{A_0B_t}\) is required to be Hermitian. For integers \(n\geq2\),
the generalized R\'enyi entropy is defined as
\begin{equation}
    S_n
    =
    \frac{1}{1-n}\log\tr\!\left(T_{A_0B_t}^{\,n}\right).
    \label{eq:def}
\end{equation}

Once an appropriate branch prescription for the logarithm is specified,
the corresponding von Neumann entropy can be defined as
\begin{equation}
    S
    =
    -\tr\!\left(T_{A_0B_t}\log T_{A_0B_t}\right).
    \label{vN_definition}
\end{equation}

The construction above provides a natural framework for extending
methods developed for spacelike entanglement to timelike-separated
subsystems. In particular, the Euclidean twist-operator construction
admits a natural extension to Lorentzian signature through the real-time
replica method developed in
\cite{Dong:2016hjy,Colin-Ellerin:2020mva,Colin-Ellerin:2021jev}. The
resulting framework provides two complementary descriptions of timelike
entanglement: Lorentzian twist-operator correlators and moments of
reduced spacetime density matrices
\cite{Milekhin:2025ycm,Guo:2025dtq}. Establishing a precise correspondence
between these descriptions requires independent calculations of the
microscopic operator moments and the continuum replica correlators,
including their complex phases. The problem becomes particularly
nontrivial in the presence of physical boundaries or a mass scale, where
the correlators depend on additional boundary and dynamical data. A
systematic treatment therefore requires methods for computing the
generally non-Hermitian reduced operators and relating their spectral
properties to the Lorentzian replica construction. Such a framework also
makes it possible to identify what controls the imaginary part of
timelike entanglement and how it is constrained by causality.

Here we develop this framework for a free real scalar field. We construct
the spacetime density matrices directly on the lattice and extend
Gaussian diagonalization to determine their spectra and integer moments.
In parallel, we evaluate the corresponding Lorentzian twist correlators,
with their operator ordering fixed by the real-time replica construction.
The two approaches yield mutually consistent results for a circle and
for Dirichlet and Neumann boundaries. We further extend the comparison
beyond conformal symmetry using massive form-factor expansions. The
operator framework also yields analytic constraints on the complex
phases. In the massless Dirichlet theory, we establish a causally
connected time window in which all integer moments are real for arbitrary
initial states and strictly positive for Gaussian states. This accounts
for the real R\'enyi entropies in that window and demonstrates explicitly
that causal contact is not sufficient for a nonzero imaginary part.

\emph{Timelike entanglement and real-time replicas---}
We consider two-dimensional QFTs, with or without physical boundaries,
and spatial intervals \(A_0\subset C_0\) and \(B_t\subset C_t\) on the
Cauchy surfaces \(C_0\) and \(C_t\), respectively. The reduced spacetime
density matrix \(T_{A_0B_t}\) admits an SK path-integral representation
with cuts along \(A_0\) and \(B_t\) on the forward branch of the contour.
Replicating the contour \(n\) times and cyclically gluing the copies
along these cuts produces \(\tr (T_{A_0B_t}^{\,n})\). In two-dimensional
QFTs, the resulting replica branch points are represented by twist
and antitwist operators. The real-time replica method therefore relates
\(\tr (T_{A_0B_t}^{\,n})\) directly to Lorentzian twist-operator correlation functions \cite{Dong:2016hjy,Colin-Ellerin:2020mva,Colin-Ellerin:2021jev,Milekhin:2025ycm,Gong:2025pnu}.

For a circle of circumference \(L\), let
\(A_0=[a,b]\) and \(B_t=[c,d]\). Since the circle has no physical
boundary, all four endpoints of the two intervals are replica branch
points. The corresponding moment is therefore represented by the
Lorentzian four-point function
\begin{equation}
    \tr\!\left( T_{A_0B_t}^{\,n} \right)
    =
    \left\langle
    \sigma_n(a,0)\widetilde{\sigma}_n(b,0)
    \sigma_n(c,t)\widetilde{\sigma}_n(d,t)
    \right\rangle_{\mathrm{cyl}}.
    \label{cylinde_four_point}
\end{equation}
The operator ordering in Eq.~\eqref{cylinde_four_point}, and hence the
Lorentzian Riemann sheet on which the correlator is evaluated, is fixed
by the SK \(\mathrm{i}0^+\) prescription. Lorentzian twist correlators
for related time-separated double-interval configurations have also
been studied in free Dirac and holographic CFTs
\cite{Kawamoto:2025oko,Harper:2025lav}.

For a strip of width \(L\), we instead take
\(A_0=[0,a]\) and \(B_t=[a,L]\). The endpoints \(x=0\) and \(x=L\)
are physical boundaries rather than replica branch points. The only
branch points are therefore located at \((a,0)\) and \((a,t)\), so the
moment reduces to the BCFT two-point function
\begin{equation}
    \tr\!\left( T_{A_0B_t}^{\,n} \right)
    =
    \left\langle
    \sigma_n(a,0)\widetilde{\sigma}_n(a,t)
    \right\rangle_{\mathrm{strip}}.
    \label{strip_two_point}
\end{equation}

\begin{figure}[t]
    \centering
    \includegraphics[width=\linewidth]{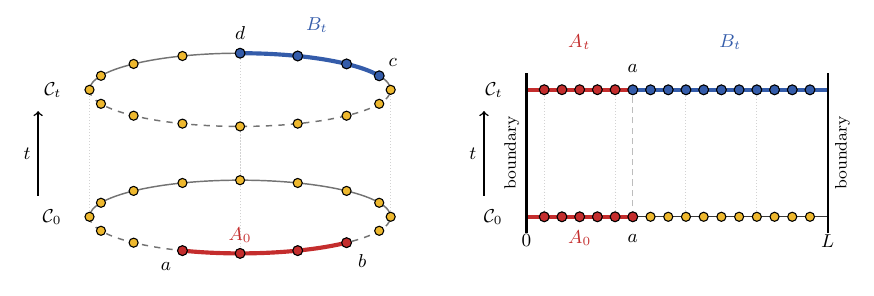}
    \caption{Schematic illustration of the reduced spacetime density
    matrix and the retained subregions in the circular (left) and strip
    (right) geometries.}
    \label{STDM}
\end{figure}

These timelike geometries probe more information than the usual
single-interval entropy in a two-dimensional CFT, for which the
twist-field two-point function is fixed by conformal invariance up to
normalization and governed by the central charge
\cite{Calabrese:2004eu,Calabrese:2009qy}. On the circle, the four-point
function in Eq.~\eqref{cylinde_four_point} depends on conformal cross
ratios and on the operator content. On the strip, the BCFT two-point
function in Eq.~\eqref{strip_two_point}, although it involves only two
bulk twist operators, contains a boundary-dependent function of the
boundary cross ratio. Numerically evaluating these correlators therefore
requires input beyond the universal single-interval result.

\section{Lattice realization}

We consider a free real scalar field on a finite spatial segment and discretize it into a lattice of coupled harmonic oscillators. The resulting Hamiltonian is
\begin{equation}
	H
	=
	\frac12 P^T P
	+
	\frac12 Q^T K Q .
\end{equation}

The coupling matrix \(K\) is determined by the mass of the scalar theory and the boundary conditions, which we take to be periodic, Dirichlet--Dirichlet (DD), Neumann--Neumann (NN), or Dirichlet--Neumann (DN).

As illustrated in Fig.~\ref{STDM}, we identify the spatial subregions
\(A_0\subset C_0\) and \(B_t\subset C_t\) with the corresponding subsets
of lattice oscillators. Keeping these oscillators and tracing out their
complements yields the reduced spacetime density matrix
\(T_{A_0B_t}\). Its R\'enyi and von Neumann entropies therefore provide a
lattice realization of timelike entanglement between the two subregions.

For the vacuum of this quadratic Hamiltonian, \(T_{A_0B_t}\) is a
generally non-Hermitian operator with a complex Gaussian kernel.
We extend standard diagonalization methods for Gaussian density
matrices~\cite{Srednicki:1993im,Peschel_1999,Katsinis:2024gef} to this
setting to obtain its complete nonzero spectrum. In terms of the
combined oscillator coordinates
\(\mathcal Q:=(q_{A_0}^{T},q_{B_t}^{T},
q_{A_0}^{\prime T},q_{B_t}^{\prime T})^{T}\),
the coordinate-basis matrix element takes the form
\begin{equation}
	\langle q_{A_0},q_{B_t}|T_{A_0B_t}|q'_{A_0},q'_{B_t}\rangle
	\propto
	\exp\left(
	-\frac12\mathcal Q^T\mathbb M\mathcal Q
	\right).
\end{equation}
We decompose the quadratic-form matrix as
\begin{equation}
	\mathbb M
	=
	\begin{pmatrix}
		L & X\\
		X^T & R
	\end{pmatrix},
	\qquad
	E:=L+R,
\end{equation}
and introduce the polynomial
\begin{equation}
	\mathcal P(z):=\det\!\left(Xz^2+Ez+X^T\right).
\end{equation}
Its roots occur in reciprocal pairs. We denote the \(d_{AB}\) roots
inside the unit circle by \(\{\xi_\mu\}_{\mu=1}^{d_{AB}}\), where
\(d_{AB}\) is the total number of lattice oscillators retained in
\(A_0\) and \(B_t\). The complete nonzero spectrum of \(T_{A_0B_t}\)
is then 
\begin{equation}
	\operatorname{spec}(T_{A_0B_t})\setminus\{0\}
	=
	\left\{
	\prod_{\mu=1}^{d_{AB}}
	(1-\xi_\mu)\xi_\mu^{m_\mu}
	\;\middle|\;
	\boldsymbol m\in\mathbb Z_{\geq0}^{d_{AB}}
	\right\},
    \label{eq:spectrum}
\end{equation}
where \(\boldsymbol m=(m_1,\ldots,m_{d_{AB}})\) runs over all
\(d_{AB}\)-tuples of nonnegative integers. The R\'enyi and von Neumann
entropies then follow directly from this spectrum through
Eqs.~\eqref{eq:def} and~\eqref{vN_definition}, respectively. Details of the
derivation are given in the Supplemental Material~\cite{SMdetail}.

\section{Consistency checks}
\begin{figure}[t]
	\centering
	\hspace{-0.49cm}
	\includegraphics[width=0.922\columnwidth]{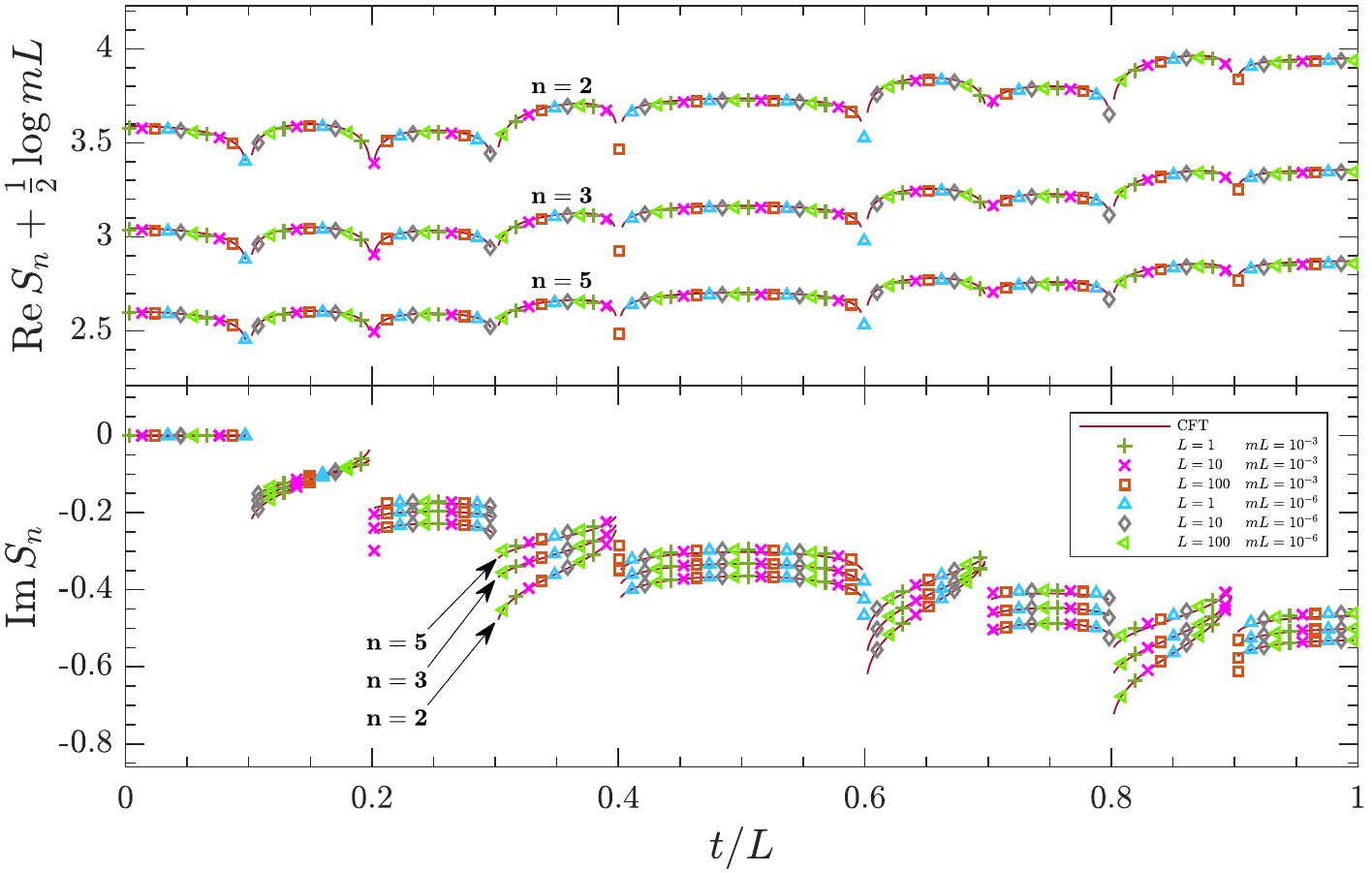}
    \caption{R\'enyi entropies for periodic boundary condition as
    functions of \(t/L\), with \(N=5000\) lattice sites in total and
    \((a,b,c,d)/L=(0.4,0.5,0.6,0.8)\). An additive shift is applied to the real part to remove the universal divergent term.}
    \label{Fig_period}
\end{figure}

\emph{Periodic boundary condition---}
Let \(A_0=[a,b]\) and \(B_t=[c,d]\) be intervals on a circle. Applying
Eq.~\eqref{cylinde_four_point} together with the result of
\cite{Calabrese:2009ez} gives the corresponding continuum
prediction, which agrees well with the lattice results, as shown in
Fig.~\ref{Fig_period}. For \(t<c-b\), the imaginary part vanishes
because \(A_0\) and \(B_t\) are causally disconnected, and the entropy
reduces to the real spacelike result. At \(t=c-b\), the endpoints
\((b,0)\) and \((c,t)\) become null separated, and the real part develops
the expected divergence described by Eq.~\eqref{cylinde_four_point}.
Similar light-cone singularities for time-separated double intervals
were identified in \cite{Kawamoto:2025oko}. For \(t>c-b\), the intervals
become causally connected and the entropy becomes complex. More generally,
both the real and imaginary parts vary sharply whenever a pair of
endpoints becomes null separated, reflecting the sensitivity of the
entropy to the causal relation between the subsystems.

\begin{figure*}[t]
	\centering
	
	\begin{minipage}[t]{\columnwidth}
		\centering
		\makebox[\linewidth][c]{%
			\hspace*{-0.20cm}%
			\includegraphics[width=0.92\linewidth]
			{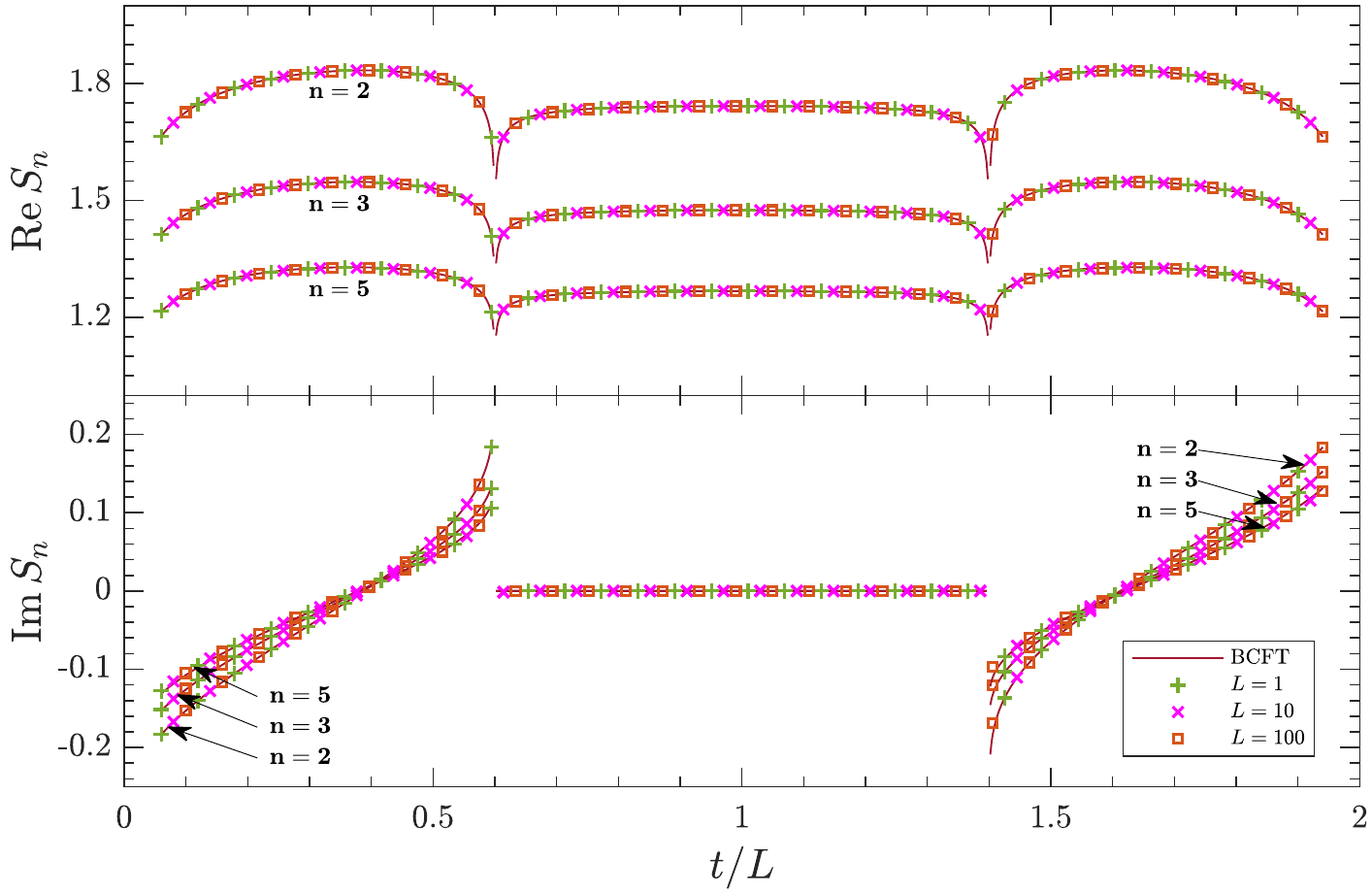}%
			\hspace*{0.20cm}%
		}
		
		\vspace{2mm}
	\end{minipage}
	\hfill
	\begin{minipage}[t]{\columnwidth}
		\centering
		\makebox[\linewidth][c]{%
			\hspace*{-0.19cm}%
			\includegraphics[width=0.92\linewidth]
			{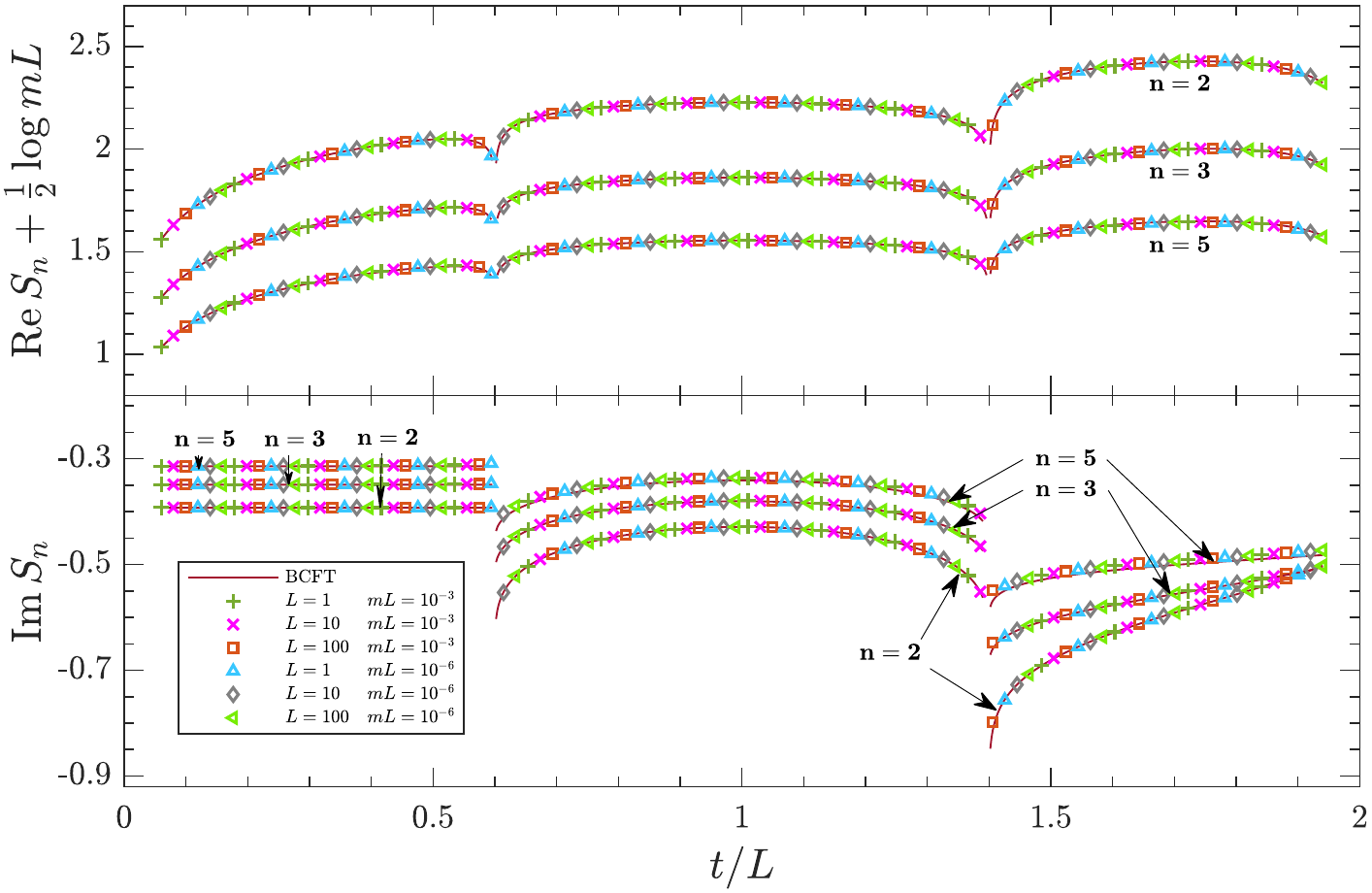}%
			\hspace*{0.19cm}%
		}
		
		\vspace{2mm}
	\end{minipage}
	
    \caption{R\'enyi entropies for DD (left) and NN (right) boundary conditions as functions of \(t/L\), with \(N=5000\) lattice sites in total and \(a/L=0.3\). An additive shift is applied to the real part of the NN result to remove the universal divergent term.}
	\label{fig:DD_NN}
\end{figure*}

\emph{Dirichlet and Neumann boundary conditions---}
Let \(A_0=[0,a]\) and \(B_t=[a,L]\) be intervals on a strip. Applying Eq.~\eqref{strip_two_point} together with the results in
\cite{Estienne:2023ekf} gives the corresponding continuum prediction, which
agrees well with the lattice results, as shown in Fig.~\ref{fig:DD_NN}.

The interior singularities at
\(t=2a\) and \(t=2L-2a\) correspond to reflected light cones, where the real part diverges while the imaginary part has finite one-sided limits.
For the NN boundary condition, in the interval \(0<t<2a\), the imaginary
part takes the constant value
\begin{equation}
	\operatorname{Im}S_{n,\mathrm{NN}}
	=
	-\frac{\pi(n+1)}{12n},
	\qquad n=2,3,\ldots .
\end{equation}
For DD boundary conditions, we find that
\begin{equation}
    \operatorname{Im}S_{n,\mathrm{DD}}=0,
    \qquad n=2,3,\ldots,
    \label{eq:realwindow}
\end{equation}
throughout \(2a<t<2L-2a\), even though the retained regions are
causally connected. We establish this result below using the same-region reduction and the partial-transpose identity.

\begin{figure}[b]
    \centering
    \makebox[\columnwidth][c]{%
        \hspace*{-0.18cm}%
        \includegraphics[width=0.933\columnwidth]{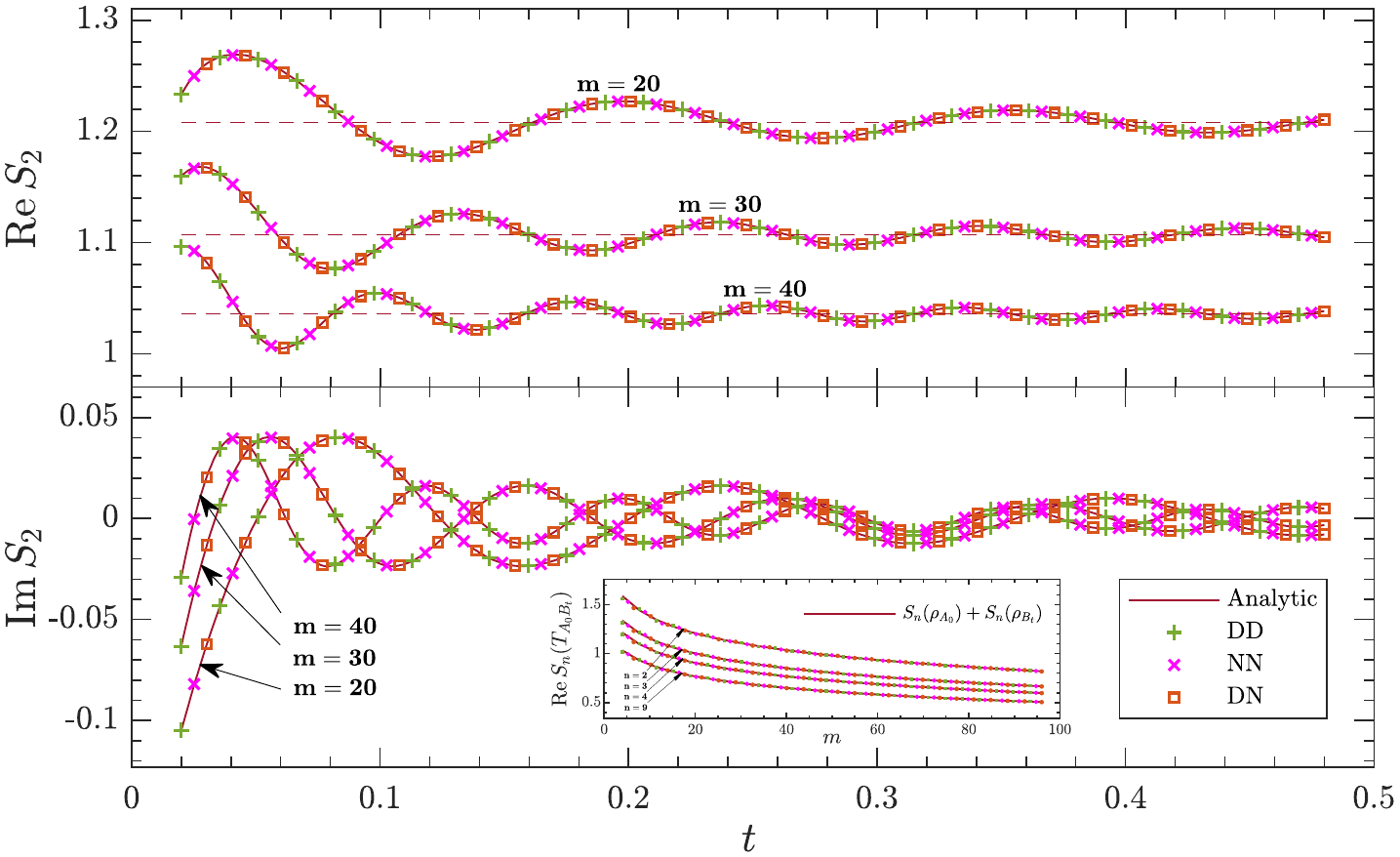}%
        \hspace*{0.18cm}%
    }
    \caption{The main panels show the R\'enyi entropies at several masses for DD,
	NN, and DN boundary conditions. The dashed lines denote the reference
	values \(S_n(\rho_{A_0})+S_n(\rho_{B_t})\). The inset shows the mass
	dependence of \(\operatorname{Re}S_n(T_{A_0B_t})\) and
	\(S_n(\rho_{A_0})+S_n(\rho_{B_t})\) at \(t=0.45\). The calculations
	use \(N=5000\) lattice sites in total, \(L=1\), and \(a=0.3\).
}
    \label{fig:massive}
\end{figure}

\emph{Beyond conformal symmetry---}
The spacetime density matrix formalism also applies to QFTs without
conformal symmetry. In massive integrable QFTs, form-factor methods for
branch-point twist fields provide a systematic framework for studying
entanglement, including recent studies of timelike entanglement in
massive scalar theories
\cite{Castro-Alvaredo:2026ohy}. For a massive scalar field on the full
line, the twist-field two-point function admits a form-factor cumulant
expansion~\cite{Bianchini:2016mra}. At sufficiently large mass \(m\) and
before the first reflected light-cone singularity, \(0<t<2a\), boundary
effects are expected to be suppressed. The finite-segment result is
therefore well approximated by the full-line result. The SK continuation
then gives the corresponding continuum prediction, which agrees well
with the lattice results for all three boundary conditions, as shown in
Fig.~\ref{fig:massive}.

In massive QFTs, the mass scale is known to control the large-distance
behavior of spacelike entanglement, with correlation length
\(\xi\sim m^{-1}\)~\cite{Cardy:2007mb}. Interestingly, we find that the
same scale also governs the timelike behavior. As shown in
Fig.~\ref{fig:massive}, once \(t\gg\xi\), the timelike R\'enyi entropy
approaches the sum of the corresponding spacelike R\'enyi entropies and
becomes real. This behavior differs from that of the spacelike R\'enyi
entropy, which, for an interval with \(\ell\gg\xi\), saturates at a value
controlled by the correlation length, \(S_n\sim\log\xi\). At fixed time,
this tendency becomes more pronounced as the mass increases, providing
numerical evidence for the approximate factorization
\(T_{A_0B_t}\approx\rho_{A_0}\otimes\rho_{B_t}\) \cite{Milekhin:2025ycm}.


\emph{Reality window---}
The DD reality window in Eq.~\eqref{eq:realwindow} follows from the
same-region setup. Let \(A_0=A_t=[0,a]\) and
\(B_t=[a,L]\) be intervals on a strip with Dirichlet boundary
condition. For a massless real scalar field, in the window
\(2a<t<2L-2a\), the direct and reflected contributions to the
Pauli--Jordan function between \(A_0\) and \(A_t\) cancel. Consequently,
the Weyl operators associated with these two regions commute, and so do
their local operator algebras. It then follows that \(T_{A_0A_t}\) is positive semidefinite: its trace pairing with every positive operator on the tensor-product Hilbert space is nonnegative, by the commutativity of the two local algebras and the positive semidefiniteness of \(\rho_0\).

The conclusions below extend to mixed initial states by purification
with an auxiliary reference system. For simplicity, we first consider a
pure initial state and use the complementary-region identity
\cite{Milekhin:2025ycm}:
\begin{equation}
    \tr\!\left(T_{A_0B_t}^{\,n}\right)
    =
    \tr\!\left[
        \left(T_{A_0A_t}^{T_{A_t}}\right)^n
    \right],
    \label{eq:main-complementary-moments}
\end{equation}
for integers \(n\geq1\), where the superscript \(T_{A_t}\) denotes partial transposition on the later-time factor. These partial-transpose moments also appear in replica calculations of entanglement negativity for reduced density matrices~\cite{Calabrese:2012ew}. The identity therefore suggests a
possible connection between timelike R\'enyi entropies and a generalized
notion of entanglement negativity.

Since \(T_{A_0A_t}\) is positive semidefinite, it is Hermitian, and
its partial transpose is therefore Hermitian with a real spectrum.
Equation~\eqref{eq:main-complementary-moments} then implies that, for
\emph{any} initial state, all moments of \(T_{A_0B_t}\) are real. In particular, the even moments are strictly positive, being sums of even powers of real eigenvalues, and cannot vanish because \(\tr \!\left(T_{A_0B_t}\right)=1\). The odd moments, by contrast, may be negative or vanish.

For Gaussian initial states, pure or mixed, the preceding constraints can
be strengthened. Partial transposition preserves the Gaussian form, so
the spectral construction developed for \(T_{A_0A_t}\) applies equally
to \(T_{A_0A_t}^{T_{A_t}}\). Its nonzero spectrum has the form in
Eq.~\eqref{eq:spectrum}, with real parameters
\(\widetilde{\xi}_\mu\) satisfying
\(\lvert\widetilde{\xi}_\mu\rvert<1\). Consequently, all integer moments
are strictly positive, and all integer R\'enyi entropies are real on the
positive-real logarithmic branch.

Although \(A_0\) and \(B_t\) remain causally connected throughout this
window, all moments are strictly positive for initial Gaussian states.
Thus, causal contact alone does not guarantee a nonzero imaginary
part of the R\'enyi entropy on the chosen branch. The same-region
reduction reveals additional constraints on the moments imposed
by dynamics and boundary conditions.

\section{Discussion}

We have developed a systematic Gaussian method for determining the
spectra and integer moments of non-Hermitian reduced spacetime
density matrices in a free scalar theory, making timelike entanglement directly computable
from a microscopic operator construction. Independently evaluated
Lorentzian twist correlators provide a complementary continuum
description. Their agreement in both magnitude and phase across
periodic, boundary, and massive settings establishes a consistent
connection between the two constructions and extends explicit
calculations beyond universal conformal results.
The framework also yields analytic insight, as illustrated by the
Dirichlet reality window: constraints on the moment phases explain
why causal contact alone does not guarantee a nonzero imaginary
R\'enyi entropy. Extending these methods to more general states
and interacting systems would broaden their applicability.
These results open new possibilities for using entanglement beyond
equal-time slices to probe quantum dynamics and spacetime correlations
in many-body systems.


~\\

\begin{acknowledgments}
The authors thank Olalla A. Castro-Alvaredo, René Meyer, Fabio Ori, Alexandre Serantes, Tadashi Takayanagi and  Jiaju Zhang for useful discussions.  The authors used ChatGPT to assist with code generation and to check intermediate steps in the analytical derivations.  This work was supported by the National Natural Science Foundation of China under Grant No.~12675068 and the Hubei Provincial Natural Science Foundation of China under Grant No.~2025AFB557. WZG also thanks the support from the Seventh Young Faculty Development Program of Huazhong University of Science and Technology. 
\end{acknowledgments}

\bibliography{references}

@article{Amico:2007ag,
    author = "Amico, Luigi and Fazio, Rosario and Osterloh, Andreas and Vedral, Vlatko",
    title = "{Entanglement in many-body systems}",
    eprint = "quant-ph/0703044",
    archivePrefix = "arXiv",
    doi = "10.1103/RevModPhys.80.517",
    journal = "Rev. Mod. Phys.",
    volume = "80",
    pages = "517--576",
    year = "2008"
}

@article{Calabrese:2009qy,
    author = "Calabrese, Pasquale and Cardy, John",
    title = "{Entanglement entropy and conformal field theory}",
    eprint = "0905.4013",
    archivePrefix = "arXiv",
    primaryClass = "cond-mat.stat-mech",
    doi = "10.1088/1751-8113/42/50/504005",
    journal = "J. Phys. A",
    volume = "42",
    pages = "504005",
    year = "2009"
}

@book{Zeng:2015pxf,
    author = "Zeng, Bei and Chen, Xie and Zhou, Duan-Lu and Wen, Xiao-Gang",
    title = "{Quantum Information Meets Quantum Matter: From Quantum Entanglement to Topological Phases of Many-Body Systems}",
    eprint = "1508.02595",
    archivePrefix = "arXiv",
    primaryClass = "cond-mat.str-el",
    doi = "10.1007/978-1-4939-9084-9",
    isbn = "978-1-4939-9082-5, 978-1-4939-9084-9",
    publisher = "Springer",
    series = "Quantum Science and Technology",
    year = "2019"
}

@book{Rangamani:2016dms,
    author = "Rangamani, Mukund and Takayanagi, Tadashi",
    title = "{Holographic Entanglement Entropy}",
    eprint = "1609.01287",
    archivePrefix = "arXiv",
    primaryClass = "hep-th",
    reportNumber = "YITP-16-106, YITP-16-106",
    doi = "10.1007/978-3-319-52573-0",
    publisher = "Springer",
    volume = "931",
    year = "2017"
}

@article{Calabrese:2004eu,
    author = "Calabrese, Pasquale and Cardy, John L.",
    title = "{Entanglement entropy and quantum field theory}",
    eprint = "hep-th/0405152",
    archivePrefix = "arXiv",
    doi = "10.1088/1742-5468/2004/06/P06002",
    journal = "J. Stat. Mech.",
    volume = "0406",
    pages = "P06002",
    year = "2004"
}

@article{Bombelli:1986rw,
    author = "Bombelli, Luca and Koul, Rabinder K. and Lee, Joohan and Sorkin, Rafael D.",
    title = "{A Quantum Source of Entropy for Black Holes}",
    reportNumber = "PRINT-86-0371 (SYRACUSE)",
    doi = "10.1103/PhysRevD.34.373",
    journal = "Phys. Rev. D",
    volume = "34",
    pages = "373--383",
    year = "1986"
}

@article{Srednicki:1993im,
    author = "Srednicki, Mark",
    title = "{Entropy and area}",
    eprint = "hep-th/9303048",
    archivePrefix = "arXiv",
    reportNumber = "LBL-33754, CFPA-93-02",
    doi = "10.1103/PhysRevLett.71.666",
    journal = "Phys. Rev. Lett.",
    volume = "71",
    pages = "666--669",
    year = "1993"
}

@article{Callan:1994py,
    author = "Callan, Jr., Curtis G. and Wilczek, Frank",
    title = "{On geometric entropy}",
    eprint = "hep-th/9401072",
    archivePrefix = "arXiv",
    reportNumber = "IASSNS-HEP-93-87",
    doi = "10.1016/0370-2693(94)91007-3",
    journal = "Phys. Lett. B",
    volume = "333",
    pages = "55--61",
    year = "1994"
}

@article{Vidal:2002rm,
    author = "Vidal, G. and Latorre, J. I. and Rico, E. and Kitaev, A.",
    title = "{Entanglement in quantum critical phenomena}",
    eprint = "quant-ph/0211074",
    archivePrefix = "arXiv",
    doi = "10.1103/PhysRevLett.90.227902",
    journal = "Phys. Rev. Lett.",
    volume = "90",
    pages = "227902",
    year = "2003"
}

@article{peschel2003calculation,
  title={Calculation of reduced density matrices from correlation functions},
  author={Peschel, Ingo},
  journal={Journal of Physics A: Mathematical and General},
  volume={36},
  number={14},
  pages={L205--L208},
  year={2003}
}

@article{Casini:2009sr,
    author = "Casini, H. and Huerta, M.",
    title = "{Entanglement entropy in free quantum field theory}",
    eprint = "0905.2562",
    archivePrefix = "arXiv",
    primaryClass = "hep-th",
    doi = "10.1088/1751-8113/42/50/504007",
    journal = "J. Phys. A",
    volume = "42",
    pages = "504007",
    year = "2009"
}

@article{aharonov1964time,
  title={Time symmetry in the quantum process of measurement},
  author={Aharonov, Yakir and Bergmann, Peter G and Lebowitz, Joel L},
  journal={Physical Review},
  volume={134},
  number={6B},
  pages={B1410},
  year={1964},
  publisher={APS}
}

@misc{Aharonov:2007czt,
    author = "Aharonov, Yakir and Vaidman, Lev",
    title = "{The Two-State Vector Formalism of Qauntum Mechanics: an Updated Review}",
    eprint = "quant-ph/0105101",
    archivePrefix = "arXiv",
    month = "6",
    year = "2007"
}

@misc{Marcovitch:2011vfd,
    author = "Marcovitch, Shmuel and Reznik, Benni",
    title = "{Structural unification of space and time correlations in quantum theory}",
    eprint = "1103.2557",
    archivePrefix = "arXiv",
    primaryClass = "quant-ph",
    month = "3",
    year = "2011"
}

@article{Fitzsimons:2013gga,
    author = "Fitzsimons, Joseph F. and Jones, Jonathan A. and Vedral, Vlatko",
    title = "{Quantum correlations which imply causation}",
    eprint = "1302.2731",
    archivePrefix = "arXiv",
    primaryClass = "quant-ph",
    doi = "10.1038/srep18281",
    journal = "Sci. Rep.",
    volume = "5",
    number = "1",
    pages = "18281",
    year = "2015"
}

@misc{Buscemi:2013xlk,
    author = "Buscemi, Francesco and Dall'Arno, Michele and Ozawa, Masanao and Vedral, Vlatko",
    title = "{Direct observation of any two-point quantum correlation function}",
    eprint = "1312.4240",
    archivePrefix = "arXiv",
    primaryClass = "quant-ph",
    month = "12",
    year = "2013"
}

@article{Leifer:2013ghr,
    author = "Leifer, M. S. and Spekkens, Robert W.",
    title = "{Towards a formulation of quantum theory as a causally neutral theory of Bayesian inference}",
    eprint = "1107.5849",
    archivePrefix = "arXiv",
    primaryClass = "quant-ph",
    doi = "10.1103/PhysRevA.88.052130",
    journal = "Phys. Rev. A",
    volume = "88",
    number = "5",
    pages = "052130",
    year = "2013"
}

@misc{Horsman:2017nqa,
    author = "Horsman, Dominic and Heunen, Chris and Pusey, Matthew F. and Barrett, Jonathan and Spekkens, Robert W.",
    title = "{Can a quantum state over time resemble a quantum state at a single time?}",
    eprint = "1607.03637",
    archivePrefix = "arXiv",
    primaryClass = "quant-ph",
    doi = "10.1098/rspa.2017.0395",
    month = "9",
    year = "2017"
}

@article{Cotler:2017anu,
    author = "Cotler, Jordan and Jian, Chao-Ming and Qi, Xiao-Liang and Wilczek, Frank",
    title = "{Superdensity Operators for Spacetime Quantum Mechanics}",
    eprint = "1711.03119",
    archivePrefix = "arXiv",
    primaryClass = "quant-ph",
    doi = "10.1007/JHEP09(2018)093",
    journal = "JHEP",
    volume = "09",
    pages = "093",
    year = "2018"
}

@article{Fullwood:2022rjd,
    author = "Fullwood, James and Parzygnat, Arthur J.",
    title = "{On quantum states over time}",
    eprint = "2202.03607",
    archivePrefix = "arXiv",
    primaryClass = "quant-ph",
    doi = "10.1098/rspa.2022.0104",
    journal = "Proc. Roy. Soc. Lond. A",
    volume = "478",
    number = "2264",
    pages = "20220104",
    year = "2022"
}

@article{Parzygnat:2022pax,
    author = "Parzygnat, Arthur J. and Fullwood, James",
    title = "{From Time-Reversal Symmetry to Quantum Bayes{\textquoteright} Rules}",
    eprint = "2212.08088",
    archivePrefix = "arXiv",
    primaryClass = "quant-ph",
    doi = "10.1103/PRXQuantum.4.020334",
    journal = "PRX Quantum",
    volume = "4",
    number = "2",
    pages = "020334",
    year = "2023"
}

@article{Lie:2024kbl,
    author = "Lie, Seok Hyung and Fullwood, James",
    title = "{Multipartite Quantum States over Time from Two Fundamental Assumptions}",
    eprint = "2410.22630",
    archivePrefix = "arXiv",
    primaryClass = "quant-ph",
    doi = "10.1103/lbf3-snp8",
    journal = "Phys. Rev. Lett.",
    volume = "135",
    number = "23",
    pages = "230204",
    year = "2025"
}

@article{Diaz:2020dfe,
    author = "Diaz, N. L. and Matera, J. M. and Rossignoli, R.",
    title = "{Spacetime Quantum Actions}",
    eprint = "2010.09136",
    archivePrefix = "arXiv",
    primaryClass = "quant-ph",
    doi = "10.1103/PhysRevD.103.065011",
    journal = "Phys. Rev. D",
    volume = "103",
    number = "6",
    pages = "065011",
    year = "2021"
}

@misc{Milekhin:2025ycm,
    author = "Milekhin, Alexey and Adamska, Zofia and Preskill, John",
    title = "{Observable and computable entanglement in time}",
    eprint = "2502.12240",
    archivePrefix = "arXiv",
    primaryClass = "quant-ph",
    month = "2",
    year = "2025"
}

@article{Guo:2025dtq,
    author = "Guo, Wu-zhong",
    title = "{Spacetime density matrix: formalism and properties}",
    eprint = "2508.20397",
    archivePrefix = "arXiv",
    primaryClass = "hep-th",
    doi = "10.1007/JHEP01(2026)128",
    journal = "JHEP",
    volume = "01",
    pages = "128",
    year = "2026"
}

@article{Das:2025fcd,
    author = "Das, Rathindra Nath and Kundu, Arnab and Martins Costa, Matheus H. and Sarkar, Nemai Chandra",
    title = "{Temporal correlations and chaos from spacetime kernels}",
    eprint = "2512.06078",
    archivePrefix = "arXiv",
    primaryClass = "hep-th",
    doi = "10.1007/JHEP04(2026)141",
    journal = "JHEP",
    volume = "04",
    pages = "141",
    year = "2026"
}

@misc{Diaz:2026qrf,
    author = "Diaz, N. L. and Cerezo, M. and Braccia, Paolo",
    title = "{Unifying spacetime approaches to quantum mechanics}",
    eprint = "2606.12539",
    archivePrefix = "arXiv",
    primaryClass = "quant-ph",
    reportNumber = "LA-UR-26-24636",
    month = "6",
    year = "2026"
}

@misc{Brukner:2004egd,
    author = "Brukner, Caslav and Taylor, Samuel and Cheung, Sancho and Vedral, Vlatko",
    title = "{Quantum Entanglement in Time}",
    eprint = "quant-ph/0402127",
    archivePrefix = "arXiv",
    month = "2",
    year = "2004"
}

@article{Fritz:2010qzm,
    author = "Fritz, Tobias",
    title = "{Quantum correlations in the temporal Clauser{\textendash}Horne{\textendash}Shimony{\textendash}Holt (CHSH) scenario}",
    eprint = "1005.3421",
    archivePrefix = "arXiv",
    primaryClass = "quant-ph",
    doi = "10.1088/1367-2630/12/8/083055",
    journal = "New J. Phys.",
    volume = "12",
    number = "8",
    pages = "083055",
    year = "2010"
}

@article{Olson:2010jy,
    author = "Olson, S. Jay and Ralph, Timothy C.",
    title = "{Entanglement between the future and past in the quantum vacuum}",
    eprint = "1003.0720",
    archivePrefix = "arXiv",
    primaryClass = "quant-ph",
    doi = "10.1103/PhysRevLett.106.110404",
    journal = "Phys. Rev. Lett.",
    volume = "106",
    pages = "110404",
    year = "2011"
}

@article{Hastings:2014qqa,
    author = "Hastings, M. B. and Mahajan, R.",
    title = "{Connecting Entanglement in Time and Space: Improving the Folding Algorithm}",
    eprint = "1411.7950",
    archivePrefix = "arXiv",
    primaryClass = "quant-ph",
    doi = "10.1103/PhysRevA.91.032306",
    journal = "Phys. Rev. A",
    volume = "91",
    number = "3",
    pages = "032306",
    year = "2015"
}

@article{Wang:2018jva,
    author = "Wang, Peng and Wu, Houwen and Yang, Haitang",
    title = "{Fix the dual geometries of $T\bar{T}$ deformed CFT$_2$ and highly excited states of CFT$_2$}",
    eprint = "1811.07758",
    archivePrefix = "arXiv",
    primaryClass = "hep-th",
    reportNumber = "CTP-SCU/2018007",
    doi = "10.1140/epjc/s10052-020-08680-7",
    journal = "Eur. Phys. J. C",
    volume = "80",
    number = "12",
    pages = "1117",
    year = "2020"
}

@article{Lerose:2021svg,
    author = "Lerose, Alessio and Sonner, Michael and Abanin, Dmitry A.",
    title = "{Scaling of temporal entanglement in proximity to integrability}",
    eprint = "2104.07607",
    archivePrefix = "arXiv",
    primaryClass = "quant-ph",
    doi = "10.1103/PhysRevB.104.035137",
    journal = "Phys. Rev. B",
    volume = "104",
    number = "3",
    pages = "035137",
    year = "2021"
}

@article{Giudice:2021smd,
    author = "Giudice, Giacomo and Giudici, Giuliano and Sonner, Michael and Thoenniss, Julian and Lerose, Alessio and Abanin, Dmitry A. and Piroli, Lorenzo",
    title = "{Temporal Entanglement, Quasiparticles, and the Role of Interactions}",
    eprint = "2112.14264",
    archivePrefix = "arXiv",
    primaryClass = "cond-mat.stat-mech",
    doi = "10.1103/PhysRevLett.128.220401",
    journal = "Phys. Rev. Lett.",
    volume = "128",
    number = "22",
    pages = "220401",
    year = "2022"
}

@article{Foligno:2023dih,
    author = "Foligno, Alessandro and Zhou, Tianci and Bertini, Bruno",
    title = "{Temporal Entanglement in Chaotic Quantum Circuits}",
    eprint = "2302.08502",
    archivePrefix = "arXiv",
    primaryClass = "quant-ph",
    reportNumber = "MIT-CTP/5368",
    doi = "10.1103/PhysRevX.13.041008",
    journal = "Phys. Rev. X",
    volume = "13",
    number = "4",
    pages = "041008",
    year = "2023"
}

@article{Liu:2022ugc,
    author = "Liu, Bowei and Chen, Hao and Lian, Biao",
    title = "{Entanglement entropy of free fermions in timelike slices}",
    eprint = "2210.03134",
    archivePrefix = "arXiv",
    primaryClass = "cond-mat.stat-mech",
    doi = "10.1103/PhysRevB.110.144306",
    journal = "Phys. Rev. B",
    volume = "110",
    number = "14",
    pages = "144306",
    year = "2024"
}

@article{Doi:2022iyj,
    author = "Doi, Kazuki and Harper, Jonathan and Mollabashi, Ali and Takayanagi, Tadashi and Taki, Yusuke",
    title = "{Pseudoentropy in dS/CFT and Timelike Entanglement Entropy}",
    eprint = "2210.09457",
    archivePrefix = "arXiv",
    primaryClass = "hep-th",
    reportNumber = "YITP-22-121, IMPU22-0052",
    doi = "10.1103/PhysRevLett.130.031601",
    journal = "Phys. Rev. Lett.",
    volume = "130",
    number = "3",
    pages = "031601",
    year = "2023"
}

@article{Carignano:2023xbz,
    author = "Carignano, Stefano and Marim{\'o}n, Carlos Ramos and Tagliacozzo, Luca",
    title = "{Temporal entropy and the complexity of computing the expectation value of local operators after a quench}",
    eprint = "2307.11649",
    archivePrefix = "arXiv",
    primaryClass = "cond-mat.stat-mech",
    doi = "10.1103/PhysRevResearch.6.033021",
    journal = "Phys. Rev. Res.",
    volume = "6",
    number = "3",
    pages = "033021",
    year = "2024"
}

@article{Bou-Comas:2024pxf,
    author = "Bou-Comas, Aleix and Marim{\'o}n, Carlos Ramos and Schneider, Jan T. and Carignano, Stefano and Tagliacozzo, Luca",
    title = "{Measuring temporal entropies in experiments}",
    eprint = "2409.05517",
    archivePrefix = "arXiv",
    primaryClass = "quant-ph",
    doi = "10.1103/436b-cnh8",
    journal = "Phys. Rev. Res.",
    volume = "8",
    number = "2",
    pages = "023229",
    year = "2026"
}

@article{Dong:2016hjy,
    author = "Dong, Xi and Lewkowycz, Aitor and Rangamani, Mukund",
    title = "{Deriving covariant holographic entanglement}",
    eprint = "1607.07506",
    archivePrefix = "arXiv",
    primaryClass = "hep-th",
    doi = "10.1007/JHEP11(2016)028",
    journal = "JHEP",
    volume = "11",
    pages = "028",
    year = "2016"
}

@article{Colin-Ellerin:2020mva,
    author = "Colin-Ellerin, Sean and Dong, Xi and Marolf, Donald and Rangamani, Mukund and Wang, Zhencheng",
    title = "{Real-time gravitational replicas: Formalism and a variational principle}",
    eprint = "2012.00828",
    archivePrefix = "arXiv",
    primaryClass = "hep-th",
    doi = "10.1007/JHEP05(2021)117",
    journal = "JHEP",
    volume = "05",
    pages = "117",
    year = "2021"
}

@article{Colin-Ellerin:2021jev,
    author = "Colin-Ellerin, Sean and Dong, Xi and Marolf, Donald and Rangamani, Mukund and Wang, Zhencheng",
    title = "{Real-time gravitational replicas: low dimensional examples}",
    eprint = "2105.07002",
    archivePrefix = "arXiv",
    primaryClass = "hep-th",
    doi = "10.1007/JHEP08(2021)171",
    journal = "JHEP",
    volume = "08",
    pages = "171",
    year = "2021"
}

@article{Gong:2025pnu,
    author = "Gong, XiangKun and Guo, Wu-zhong and Xu, Jin",
    title = "{Entanglement measures for causally connected subregions and holography}",
    eprint = "2508.05158",
    archivePrefix = "arXiv",
    primaryClass = "hep-th",
    doi = "10.1103/771p-4rkf",
    journal = "Phys. Rev. D",
    volume = "113",
    number = "10",
    pages = "106009",
    year = "2026"
}

@article{Calabrese:2009ez,
    author = "Calabrese, Pasquale and Cardy, John and Tonni, Erik",
    title = "{Entanglement entropy of two disjoint intervals in conformal field theory}",
    eprint = "0905.2069",
    archivePrefix = "arXiv",
    primaryClass = "hep-th",
    doi = "10.1088/1742-5468/2009/11/P11001",
    journal = "J. Stat. Mech.",
    volume = "0911",
    pages = "P11001",
    year = "2009"
}

@article{Estienne:2023ekf,
    author = "Estienne, Benoit and Ikhlef, Yacine and Rotaru, Andrei and Tonni, Erik",
    title = "{Entanglement entropies of an interval for the massless scalar field in the presence of a boundary}",
    eprint = "2308.00614",
    archivePrefix = "arXiv",
    primaryClass = "hep-th",
    doi = "10.1007/JHEP05(2024)236",
    journal = "JHEP",
    volume = "05",
    pages = "236",
    year = "2024"
}

@article{Bianchini:2016mra,
    author = "Bianchini, Davide and Castro-Alvaredo, Olalla A.",
    title = "{Branch Point Twist Field Correlators in the Massive Free Boson Theory}",
    eprint = "1607.05656",
    archivePrefix = "arXiv",
    primaryClass = "hep-th",
    doi = "10.1016/j.nuclphysb.2016.10.016",
    journal = "Nucl. Phys. B",
    volume = "913",
    pages = "879--911",
    year = "2016"
}

@article{Castro-Alvaredo:2026ohy,
    author = "Castro-Alvaredo, Olalla A.",
    title = "{Temporal entanglement in quantum field theory}",
    eprint = "2603.20765",
    archivePrefix = "arXiv",
    primaryClass = "hep-th",
    doi = "10.1088/1751-8121/ae9b51",
    journal = "J. Phys. A",
    volume = "59",
    number = "35",
    pages = "355201",
    year = "2026"
}

@article{Peschel_1999,
   title={Density matrices for a chain of oscillators},
   volume={32},
   number={48},
   journal={Journal of Physics A: Mathematical and General},
   publisher={IOP Publishing},
   author={Peschel, Ingo and Chung, Ming-Chiang},
   year={1999},
   month=Nov, 
   pages={8419--8428},
   eprint = "9906224",
   archivePrefix = "arXiv",
   primaryClass = "cond-mat.stat-mech",
}

@article{Katsinis:2024gef,
    author = "Katsinis, Dimitrios and Pastras, Georgios",
    title = "{Entanglement in (1+1)-dimensional free scalar field theory: Tiptoeing between continuum and discrete formulations}",
    eprint = "2406.11031",
    archivePrefix = "arXiv",
    primaryClass = "hep-th",
    doi = "10.1103/PhysRevD.110.085015",
    journal = "Phys. Rev. D",
    volume = "110",
    number = "8",
    pages = "085015",
    year = "2024"
}

@article{GuoKuoLin2011,
	author  = {Guo, Chun-Hua and Kuo, Yueh-Cheng and Lin, Wen-Wei},
    title   = {Complex symmetric stabilizing solution of the matrix equation
              {$X+A^T X^{-1}A=Q$}},
    journal = {Linear Algebra Appl.},
	volume  = {435},
	number  = {6},
	pages   = {1187--1192},
	year    = {2011},
	doi     = {10.1016/j.laa.2011.03.034}
}

@article{Doyon:2025xvo,
    author = "Doyon, Benjamin",
    title = "{Twist Fields in Many-Body Physics}",
    eprint = "2509.15064",
    archivePrefix = "arXiv",
    primaryClass = "math-ph",
    doi = "10.3390/e27121230",
    journal = "Entropy",
    volume = "27",
    number = "12",
    pages = "1230",
    year = "2025"
}

@article{Cardy:2007mb,
    author = "Cardy, J. L. and Castro-Alvaredo, O. A. and Doyon, B.",
    title = "{Form factors of branch-point twist fields in quantum integrable models and entanglement entropy}",
    eprint = "0706.3384",
    archivePrefix = "arXiv",
    primaryClass = "hep-th",
    doi = "10.1007/s10955-007-9422-x",
    journal = "J. Statist. Phys.",
    volume = "130",
    pages = "129--168",
    year = "2008"
}

@article{Kawamoto:2025oko,
    author = "Kawamoto, Taishi and Maeda, Ryota and Nakamura, Nanami and Takayanagi, Tadashi",
    title = "{Traversable AdS wormhole via non-local double trace or Janus deformation}",
    eprint = "2502.03531",
    archivePrefix = "arXiv",
    primaryClass = "hep-th",
    reportNumber = "YITP-25-09",
    doi = "10.1007/JHEP04(2025)086",
    journal = "JHEP",
    volume = "04",
    pages = "086",
    year = "2025"
}

@article{Harper:2025lav,
    author = "Harper, Jonathan and Kawamoto, Taishi and Maeda, Ryota and Nakamura, Nanami and Takayanagi, Tadashi",
    title = "{Non-Hermitian density matrices from timelike entanglement and wormholes}",
    eprint = "2512.13800",
    archivePrefix = "arXiv",
    primaryClass = "hep-th",
    reportNumber = "YITP-25-186",
    doi = "10.1103/j2xw-tcsb",
    journal = "Phys. Rev. D",
    volume = "113",
    number = "12",
    pages = "126017",
    year = "2026"
}

@misc{SMdetail,
  note = {This Supplemental Material provides technical details supporting the main text. Section S1 describes the lattice construction of spacetime density matrices and derives the spectrum and entropies of the reduced operator. Section S2 analyzes the Dirichlet reality window and derives the associated positivity and moment constraints. Section S3 presents the Lorentzian twist-operator correlation functions and analyzes the phase structure in the strip geometry. Section S4 specifies the numerical branch prescriptions and explains how singularities are handled in the plots.}
}

@article{Sorkin:2012sn,
    author = "Sorkin, Rafael D.",
    title = "{Expressing entropy globally in terms of (4D) field-correlations}",
    eprint = "1205.2953",
    archivePrefix = "arXiv",
    primaryClass = "hep-th",
    doi = "10.1088/1742-6596/484/1/012004",
    journal = "J. Phys. Conf. Ser.",
    volume = "484",
    pages = "012004",
    year = "2014"
}

@article{Chen:2020ild,
    author = "Chen, Yangang and Hackl, Lucas and Kunjwal, Ravi and Moradi, Heidar and Yazdi, Yasaman K. and Zilh{\~a}o, Miguel",
    title = "{Towards spacetime entanglement entropy for interacting theories}",
    eprint = "2002.00966",
    archivePrefix = "arXiv",
    primaryClass = "hep-th",
    doi = "10.1007/JHEP11(2020)114",
    journal = "JHEP",
    volume = "11",
    pages = "114",
    year = "2020"
}

@article{Jones:2026vkw,
    author = "Jones, Joshua Y. L. and Yazdi, Yasaman K.",
    title = "{Spectral spacetime entropy for quasifree theories}",
    eprint = "2602.16782",
    archivePrefix = "arXiv",
    primaryClass = "hep-th",
    reportNumber = "DIAS-STP-26-04",
    doi = "10.1007/JHEP11(2026)122",
    journal = "JHEP",
    volume = "11",
    pages = "122",
    year = "2026"
}

@article{Calabrese:2012ew,
    author = "Calabrese, Pasquale and Cardy, John and Tonni, Erik",
    title = "{Entanglement negativity in quantum field theory}",
    eprint = "1206.3092",
    archivePrefix = "arXiv",
    primaryClass = "cond-mat.stat-mech",
    doi = "10.1103/PhysRevLett.109.130502",
    journal = "Phys. Rev. Lett.",
    volume = "109",
    pages = "130502",
    year = "2012"
}

\clearpage

\setlength{\textwidth}{6.4in}
\setlength{\oddsidemargin}{0.05in}
\setlength{\evensidemargin}{0.05in}

\onecolumngrid

\makeatletter
\onecolumn@grid@setup
\let\set@footnotewidth\set@footnotewidth@one
\let\compose@footnotes\compose@footnotes@one
\makeatother

\setcounter{secnumdepth}{3}

\setcounter{section}{0}
\setcounter{subsection}{0}
\setcounter{subsubsection}{0}
\setcounter{equation}{0}
\setcounter{figure}{0}
\setcounter{table}{0}
\setcounter{footnote}{0}
\setcounter{page}{1}

\renewcommand{\thesection}{S\arabic{section}}
\renewcommand{\thesubsection}{\Alph{subsection}}
\renewcommand{\thesubsubsection}{\arabic{subsubsection}}
\renewcommand{\theequation}{S\arabic{equation}}
\renewcommand{\thefigure}{S\arabic{figure}}
\renewcommand{\thetable}{S\arabic{table}}
\renewcommand{\thepage}{S\arabic{page}}

\renewcommand{\theHsection}{supp.\arabic{section}}
\renewcommand{\theHsubsection}
	{supp.\arabic{section}.\arabic{subsection}}
\renewcommand{\theHsubsubsection}
	{supp.\arabic{section}.\arabic{subsection}.\arabic{subsubsection}}
\renewcommand{\theHequation}{supp.\arabic{equation}}
\renewcommand{\theHfigure}{supp.\arabic{figure}}
\renewcommand{\theHtable}{supp.\arabic{table}}

\begin{center}
	{\large\bfseries
	Supplemental Material for\\[0.35em]
	``Timelike Entanglement from Spacetime Density Matrices:\\
	A Lattice Realization''\par}

	\vspace{0.8em}

\end{center}

\vspace{1em}

\section{Lattice Realization}
\label{sec:lattice-framework-timelike-entanglement}

\subsection{Gaussian Construction}
\label{subsec:review-spacetime-density-matrix}

\paragraph{Spacetime density matrix.}
Let \(\rho_0\) denote the initial density matrix and
\(U(t)=e^{-\mathrm{i}Ht}\) the time-evolution operator.
In an orthonormal basis \(\{|i\rangle\}\), the full spacetime density
matrix takes the explicit form
\begin{equation}
	T_{C_0C_t}
	=
	\sum_{i,j}
	|j\rangle\langle i|
	\otimes
	U\rho_0|i\rangle\langle j|U^\dagger,
	\qquad
	U:=U(t).
	\label{eq:full-spacetime-density-matrix-general}
\end{equation}

For subsystems \(A_0\subset C_0\) and \(B_t\subset C_t\), tracing out
their complements gives the reduced spacetime density matrix
\begin{equation}
	T_{A_0B_t}
	=
	\operatorname{tr}_{\bar A_0\bar B_t}T_{C_0C_t}.
	\label{eq:reduced-spacetime-density-matrix-general}
\end{equation}

In terms of its nonzero eigenvalues \(\lambda\), counted with algebraic
multiplicity, the R\'enyi entropy is
\begin{equation}
	S_n(T_{A_0B_t})
	=
	\frac{1}{1-n}\log\operatorname{tr}\!\left( T_{A_0B_t}^{\,n} \right)
	=
	\frac{1}{1-n}\log\sum\lambda^n,
	\qquad n=2,3,\ldots.
	\label{eq:renyi-definition-reduced-spacetime}
\end{equation}

If an analytic continuation in \(n\) is available near \(n=1\),
the von Neumann entropy is
\begin{equation}
	S(T_{A_0B_t})
	=
	-\left.
	\frac{\partial}{\partial n}
	\log\operatorname{tr}\!\left( T_{A_0B_t}^{\,n} \right)
	\right|_{n=1}
	=
	-\sum\lambda\log\lambda.
	\label{eq:von-neumann-definition-reduced-spacetime}
\end{equation}

\vspace{\baselineskip}
\paragraph{Oscillator setup.}

Consider \(N\) coupled harmonic oscillators with Hamiltonian
\begin{equation}
	H
	=
	\frac12 P^T P
	+
	\frac12 Q^T K Q,
	\label{eq:oscillator-hamiltonian}
\end{equation}
where \(Q=(Q_1,\ldots,Q_N)^T\) and \(P=(P_1,\ldots,P_N)^T\) collect the position and momentum operators. The coupling matrix \(K\) is fixed, real, symmetric, and positive definite. It can therefore be diagonalized by an orthogonal matrix:
\begin{equation}
	K
	=
	O\,\operatorname{diag}(\omega_1^2,\ldots,\omega_N^2)\,O^T,
	\qquad
	\omega_\alpha>0.
	\label{eq:K-diagonalization}
\end{equation}

For later convenience, we define
\begin{align}
	E_0
	&:=
	\frac{1}{2}
	\bigl(\omega_1+\cdots+\omega_N\bigr),
	\label{eq:E0-definition}
	\\[0.3em]
	\Omega
	&:=
	O\,\operatorname{diag}(\omega_1,\ldots,\omega_N)\,O^T,
	\label{eq:Omega-definition}
	\\[0.3em]
	C
	&:=
	O\,\operatorname{diag}\bigl(
	\omega_\alpha\cot(\omega_\alpha t)
	\bigr)\,O^T,
	\label{eq:C-definition}
	\\[0.3em]
	F
	&:=
	O\,\operatorname{diag}\bigl(
	\omega_\alpha\csc(\omega_\alpha t)
	\bigr)\,O^T.
	\label{eq:F-definition}
\end{align}

Taking the vacuum \(\rho_0=|0\rangle\langle0|\) as the initial state, the two matrix elements needed below are
\begin{align}
	&\langle Q|U\rho_0|Q'\rangle
	=
	e^{-\mathrm{i}E_0t}
	\left(\frac{\det\Omega}{\pi^N}\right)^{1/2}
	\exp\left[
	-\frac12 Q^T\Omega Q
	-\frac12 Q^{\prime T}\Omega Q'
	\right],
	\label{eq:vacuum-evolution-matrix-element}
	\\[0.8em]
	&\langle Q|U^\dagger|Q'\rangle
	=
	\left(\frac{\det F}{(-2\pi\mathrm{i})^N}\right)^{1/2}
	\exp\left[
	-\frac{\mathrm{i}}{2}Q^TCQ
	-\frac{\mathrm{i}}{2}Q^{\prime T}CQ'
	+\mathrm{i}Q^{\prime T}FQ
	\right].
	\label{eq:adjoint-evolution-matrix-element}
\end{align}

Here and below, \(t\) is understood as \(t+\mathrm{i}0^+\) and the branches  of the square roots are chosen according to the requirements of each case.

\paragraph{Full spacetime density matrix.}
Using Eq.~\eqref{eq:full-spacetime-density-matrix-general} and the
matrix elements above, we obtain
\begin{equation}
	\langle Q_0,Q_t|T_{C_0C_t}|Q_0',Q_t'\rangle
	=
	\langle Q_t|U\rho_0|Q_0'\rangle
	\langle Q_0|U^\dagger|Q_t'\rangle
	=
	\mathcal N_{C_0C_t}
	\exp\left(
	-\frac12\mathcal Q_C^T\mathbb M_C\mathcal Q_C
	\right),
	\label{eq:full-T-kernel-matrix-form}
\end{equation}
where
\(\mathcal Q_C:=(Q_0^T,Q_t^T,Q_0^{\prime T},Q_t^{\prime T})^T\).
The normalization factor is
\begin{equation}
	\mathcal N_{C_0C_t}
	=
	e^{-\mathrm{i}E_0t}
	\left(\frac{\det\Omega}{\pi^N}\right)^{1/2}
	\left(\frac{\det F}{(-2\pi\mathrm{i})^N}\right)^{1/2}.
	\label{eq:full-T-normalization}
\end{equation}

The quadratic-form matrix is
\begin{equation}
	\mathbb M_C
	=
	\begin{pmatrix}
		\mathrm{i}C  & 0      & 0      & -\mathrm{i}F\\
		0            & \Omega & 0      & 0\\
		0            & 0      & \Omega & 0\\
		-\mathrm{i}F & 0      & 0      & \mathrm{i}C
	\end{pmatrix}.
	\label{eq:MC-definition}
\end{equation}

\vspace{\baselineskip}
\paragraph{Reduced spacetime density matrix.}
Following Eq.~\eqref{eq:reduced-spacetime-density-matrix-general},
we write
\(Q_0=(q_{A_0}^{T},u_{\bar A_0}^{T})^{T}\),
\(Q_0'=(q_{A_0}^{\prime T},u_{\bar A_0}^{T})^{T}\),
\(Q_t=(q_{B_t}^{T},v_{\bar B_t}^{T})^{T}\), and
\(Q_t'=(q_{B_t}^{\prime T},v_{\bar B_t}^{T})^{T}\),
with the complementary coordinates identified in the bra and ket.
Evaluating the resulting complex Gaussian integral gives
\begin{equation}
	\begin{aligned}
		\langle q_{A_0},q_{B_t}|T_{A_0B_t}|q_{A_0}',q_{B_t}'\rangle
		&=
		\int
		\mathrm d^{|\bar A|}u_{\bar A_0}\,
		\mathrm d^{|\bar B|}v_{\bar B_t}\,
		\langle
		q_{A_0},u_{\bar A_0};
		q_{B_t},v_{\bar B_t}
		|T_{C_0C_t}|
		q_{A_0}',u_{\bar A_0};
		q_{B_t}',v_{\bar B_t}
		\rangle
		\\[0.3em]
		&=
		\mathcal N_{AB}
		\exp\left(
		-\frac12\mathcal Q_{AB}^{T}\mathbb M_{AB}\mathcal Q_{AB}
		\right),
	\end{aligned}
	\label{eq:reduced-T-kernel-matrix-form}
\end{equation}
where
\(\mathcal Q_{AB}:=
(q_{A_0}^{T},q_{B_t}^{T},q_{A_0}^{\prime T},q_{B_t}^{\prime T})^T\).
The normalization factor is
\begin{equation}
	\mathcal N_{AB}
	=
	\mathcal N_{C_0C_t}
	\frac{(2\pi)^{(|\bar A|+|\bar B|)/2}}
	{\sqrt{\det\mathbb A_{AB}}}.
	\label{eq:reduced-T-normalization}
\end{equation}

The quadratic-form matrix and the matrices entering its construction are%
\footnote{\raggedright
For \(G\in\{\Omega,C,F\}\), \(G_{RS}\) denotes the submatrix obtained\newline
by restricting \(G\) to rows in \(R\) and columns in \(S\).\par}
\begin{align}
	\mathbb M_{AB}
	&=
	\mathbb M_{AB}^{(0)}
	-
	\mathbb B_{AB}^{T}
	\mathbb A_{AB}^{-1}
	\mathbb B_{AB},
	\label{eq:reduced-T-matrix}
	\\[0.6em]
	\mathbb M_{AB}^{(0)}
	&=
	\begin{pmatrix}
		\mathrm{i}C_{AA} & 0 & 0 & -\mathrm{i}F_{AB}\\
		0 & \Omega_{BB} & 0 & 0\\
		0 & 0 & \Omega_{AA} & 0\\
		-\mathrm{i}F_{BA} & 0 & 0 & \mathrm{i}C_{BB}
	\end{pmatrix},
	\label{eq:MAB0-definition}
	\\[0.6em]
	\mathbb A_{AB}
	&=
	\begin{pmatrix}
		\Omega_{\bar A\bar A}+\mathrm{i}C_{\bar A\bar A}
		&
		-\mathrm{i}F_{\bar A\bar B}
		\\
		-\mathrm{i}F_{\bar B\bar A}
		&
		\Omega_{\bar B\bar B}+\mathrm{i}C_{\bar B\bar B}
	\end{pmatrix},
	\label{eq:AAB-definition}
	\\[0.6em]
	\mathbb B_{AB}
	&=
	\begin{pmatrix}
		-\mathrm{i}C_{\bar A A}
		&
		0
		&
		-\Omega_{\bar A A}
		&
		\mathrm{i}F_{\bar A B}
		\\
		\mathrm{i}F_{\bar B A}
		&
		-\Omega_{\bar B B}
		&
		0
		&
		-\mathrm{i}C_{\bar B B}
	\end{pmatrix}.
	\label{eq:BAB-definition}
\end{align}

This result applies to arbitrary retained subsets \(A_0\subset C_0\)
and \(B_t\subset C_t\).

\subsection{Spectrum and Entropies}
\label{subsec:spectrum-and-entropies}

The spectra of Hermitian Gaussian reduced density matrices have been
studied in \cite{Srednicki:1993im,Peschel_1999,Katsinis:2024gef}.
Here we extend these methods to the generally non-Hermitian Gaussian
operator \(T_{A_0B_t}\). We determine its Gaussian base eigenfunction
from a stabilizing Riccati equation and then construct a complete set
of eigenfunctions using a source generating function. The resulting
nonzero spectrum gives the associated entropies.

\paragraph{Gaussian ansatz.}
We introduce the combined coordinate vectors
\(\chi:=(q_{A_0}^{T},q_{B_t}^{T})^{T}\) and
\(\chi':=(q_{A_0}^{\prime T},q_{B_t}^{\prime T})^{T}\),
with \(d_{AB}:=|A|+|B|\).
For notational simplicity, we write
\(\mathbb M\equiv\mathbb M_{AB}\) and decompose it as
\begin{equation}
	\mathbb M
	=
	\begin{pmatrix}
		L & X\\
		X^T & R
	\end{pmatrix}.
	\label{eq:M-block-decomposition}
\end{equation}

The corresponding coordinate-basis matrix element is
\begin{equation}
	\langle\chi|T_{A_0B_t}|\chi'\rangle
	=
	\mathcal N_{AB}
	\exp\left[
	-\frac12\chi^TL\chi
	-\chi^TX\chi'
	-\frac12\chi'^TR\chi'
	\right].
	\label{eq:TAB-Gaussian-matrix-element}
\end{equation}

At the nonsingular times considered here, with the
\(\mathrm{i}0^+\) prescription understood, one can show that
\(T_{A_0B_t}\) is bounded and power compact. Its nonzero spectrum
therefore consists of isolated eigenvalues of finite algebraic
multiplicity, with zero as the only possible accumulation point.
In the generic case \(\det X\neq0\), we seek a normalizable
Gaussian right eigenfunction of the form
\begin{equation}
	\psi_0(\chi)
	:=
	\langle\chi|\psi_0\rangle
	=
	\exp\left(-\frac12\chi^T\Gamma\chi\right),
	\qquad
	\Gamma^T=\Gamma,
	\qquad
	\operatorname{Re}\Gamma>0.
	\label{eq:Gaussian-base-wavefunction}
\end{equation}

With \(\Sigma:=R+\Gamma\) and \(E:=L+R\), evaluating the complex
Gaussian integral reduces the eigenvalue problem to the algebraic
matrix Riccati equation
\begin{equation}
	\Sigma=E-X\Sigma^{-1}X^T.
	\label{eq:SigmaAB-Riccati}
\end{equation}

The present Gaussian kernel satisfies the relevant unit-circle
positivity condition, which guarantees a unique complex symmetric
stabilizing solution \cite{GuoKuoLin2011}. For this solution, define
\begin{equation}
	Z:=-X\Sigma^{-1},
	\qquad
	\rho(Z)<1,
	\label{eq:ZAB-definition}
\end{equation}
where \(\rho(Z)\) denotes the spectral radius.
The Riccati equation also implies the factorization
\begin{equation}
	\mathcal P(z)
	:=
	\det\left(Xz^2+Ez+X^T\right)
	=
	\det\Sigma\,
	\det\left(I-zZ\right)
	\det\left(zI-Z\right).
	\label{eq:PAB-ZAB-factorization}
\end{equation}

The roots of \(\mathcal P(z)\) therefore occur in reciprocal pairs.
The \(d_{AB}\) roots inside the unit circle, denoted by \(\xi_\mu\)
with \(\lvert\xi_\mu\rvert<1\), are precisely the eigenvalues of \(Z\).

\vspace{\baselineskip}
\paragraph{Eigenvalue tower.}
To construct the eigenfunctions, we introduce the generating function
\begin{equation}
	\Phi_s(\chi)
	:=
	\langle\chi|\Phi_s\rangle
	=
	\exp\left[
	-\frac12\chi^T\Gamma\chi
	+s^T\chi
	-\frac12 s^TDs
	\right],
	\label{eq:Gaussian-generating-function}
\end{equation}
where \(D=D^T\) is the unique solution of the discrete Lyapunov equation
\begin{equation}
	D-Z^TDZ=\Sigma^{-1}.
	\label{eq:DAB-Lyapunov}
\end{equation}

Direct Gaussian integration gives the closure relation
\begin{equation}
	T_{A_0B_t}|\Phi_s\rangle
	=
	\lambda_0|\Phi_{Zs}\rangle.
	\label{eq:generating-function-closure}
\end{equation}

Assuming that \(Z\) is diagonalizable, we choose \(V\) such that
\(V^{-1}ZV=\operatorname{diag}(\xi_1,\ldots,\xi_{d_{AB}})\).
Expanding \(\Phi_{Vu}\) in the source variables
\(u=(u_1,\ldots,u_{d_{AB}})^T\) generates the eigenfunctions
\begin{equation}
	\psi_{\boldsymbol m}(\chi)
    :=
	\langle\chi|\psi_{\boldsymbol m}\rangle
	=
	\left.
	\left(
	\prod_{\mu=1}^{d_{AB}}
	\frac{\partial^{m_\mu}}{\partial u_\mu^{m_\mu}}
	\right)
	\Phi_{Vu}(\chi)
	\right|_{u=0}.
	\label{eq:Gaussian-excited-eigenfunctions}
\end{equation}

These eigenfunctions have the form
\(\psi_{\boldsymbol m}=H_{\boldsymbol m}\psi_0\), where
\(H_{\boldsymbol m}\) are generalized multivariate Hermite polynomials.
The closure relation then yields
\begin{equation}
	T_{A_0B_t}|\psi_{\boldsymbol m}\rangle
	=
	\lambda_0
	\left(\prod_{\mu=1}^{d_{AB}}\xi_\mu^{m_\mu}\right)
	|\psi_{\boldsymbol m}\rangle.
	\label{eq:Gaussian-excited-eigenvalue-equation}
\end{equation}

Since \(V\) is invertible, the polynomials \(H_{\boldsymbol m}\) span
the full polynomial space, and the associated Hermite--Gaussian
eigenfunctions form a complete system in \(L^2(\mathbb R^{d_{AB}})\).
Together with the spectral properties established above, this completeness
shows that the eigenvalue tower exhausts the nonzero spectrum.
The normalization \(\operatorname{tr}T_{A_0B_t}=1\) fixes \(\lambda_0\),
giving
\begin{equation}
	\operatorname{spec}\left(T_{A_0B_t}\right)\setminus\{0\}
	=
	\left\{
	\prod_{\mu=1}^{d_{AB}}
	\left(1-\xi_\mu\right)\xi_\mu^{m_\mu}
	\;\middle|\;
	\boldsymbol m\in\mathbb Z_{\geq0}^{d_{AB}}
	\right\}.
	\label{eq:TAB-spectrum}
\end{equation}

\vspace{\baselineskip}
\paragraph{Entropies.}
For integers \(n\geq2\), substituting this spectrum into
Eq.~\eqref{eq:renyi-definition-reduced-spacetime} gives the R\'enyi entropy
\begin{equation}
	S_n(T_{A_0B_t})
	=
	\frac{1}{1-n}
	\sum_{\mu=1}^{d_{AB}}
	\left[
	n\log\left(1-\xi_\mu\right)
	-
	\log\left(1-\xi_\mu^{\,n}\right)
	\right].
	\label{eq:renyi-factorized-TAB}
\end{equation}

The von Neumann entropy follows from
Eq.~\eqref{eq:von-neumann-definition-reduced-spacetime}:
\begin{equation}
	S(T_{A_0B_t})
	=
	\sum_{\mu=1}^{d_{AB}}
	\left[
	-\log\left(1-\xi_\mu\right)
	-
	\frac{\xi_\mu}{1-\xi_\mu}\log\xi_\mu
	\right].
	\label{eq:von-neumann-entropy-final}
\end{equation}

\subsection{Discretization}
\label{subsec:lattice-discretization-scalar-field}

We discretize a free real scalar field of mass \(m\) on the interval \([0,L]\), with continuum Hamiltonian
\begin{equation}
	H
	=
	\frac12
	\int_0^L\mathrm{d}x\,
	\left[
	\pi(t,x)^2
	+
	\bigl(\partial_x\phi(t,x)\bigr)^2
	+
	m^2\phi(t,x)^2
	\right].
\end{equation}

For each boundary condition \(\alpha\), let \(\epsilon_\alpha\) and \(x_j^{(\alpha)}\) denote the corresponding lattice spacing and site positions. We define
\begin{equation}
	Q_j
	=
	\sqrt{\epsilon_\alpha}\,
	\phi\!\left(t,x_j^{(\alpha)}\right),
	\qquad
	P_j
	=
	\sqrt{\epsilon_\alpha}\,
	\pi\!\left(t,x_j^{(\alpha)}\right).
\end{equation}

These variables satisfy \([Q_j,P_k]=\mathrm{i}\delta_{jk}\) and \([Q_j,Q_k]=[P_j,P_k]=0\), and the lattice Hamiltonian becomes
\begin{equation}
	H
	=
	\frac12 P^T P
	+
	\frac12 Q^T K^{(\alpha)}Q.
\end{equation}

The lattice grids for the four boundary conditions are summarized in Table~\ref{tab:lattice-grids}, where \(j=1,\ldots,N\) labels the independent lattice sites.
\begin{table}[H]
	\centering
	\caption{Boundary conditions and corresponding lattice grids.}
	\label{tab:lattice-grids}
	\small
	\renewcommand{\arraystretch}{1.55}
	\setlength{\tabcolsep}{6pt}
	\begin{tabular}{@{}clcc@{}}
		\toprule
		\(\alpha\)
		&
		boundary condition
		&
		\(\epsilon_\alpha\)
		&
		\(x_j^{(\alpha)}\)
		\\
		\midrule
		\(\mathrm{PBC}\)
		&
		\(\phi(t,x+L)=\phi(t,x)\)
		&
		\(L/N\)
		&
		\((j-1)\epsilon_{\mathrm{PBC}}\)
		\\
		\(\mathrm{DD}\)
		&
		\(\phi(t,0)=\phi(t,L)=0\)
		&
		\(L/(N+1)\)
		&
		\(j\epsilon_{\mathrm{DD}}\)
		\\
		\(\mathrm{NN}\)
		&
		\(\partial_x\phi(t,0)=\partial_x\phi(t,L)=0\)
		&
		\(L/(N-1)\)
		&
		\((j-1)\epsilon_{\mathrm{NN}}\)
		\\
		\(\mathrm{DN}\)
		&
		\(\phi(t,0)=0,\;\partial_x\phi(t,L)=0\)
		&
		\(L/N\)
		&
		\(j\epsilon_{\mathrm{DN}}\)
		\\
		\bottomrule
	\end{tabular}
\end{table}

A nearest-neighbor finite-difference discretization gives the coupling matrices listed in Table~\ref{tab:boundary-coupling-matrices}.
\begin{table}[H]
	\centering
	\caption{Coupling matrices for the four boundary conditions.}
	\label{tab:boundary-coupling-matrices}
	\small
	\renewcommand{\arraystretch}{1.8}
	\setlength{\tabcolsep}{8pt}
	\begin{tabular}{@{}cl@{}}
		\toprule
		\(\alpha\)
		&
		\multicolumn{1}{c}{\(K_{jk}^{(\alpha)}\)}
		\\
		\midrule
		\(\mathrm{PBC}\)
		&
		\(\displaystyle
		m^2\delta_{jk}
		+
		\frac{1}{\epsilon_{\mathrm{PBC}}^2}
		\left(
		2\delta_{jk}
		-
		\delta_{j,k+1}
		-
		\delta_{j,k-1}
		\right)\)
		\\
		\addlinespace[0.3em]
		\(\mathrm{DD}\)
		&
		\(\displaystyle
		m^2\delta_{jk}
		+
		\frac{1}{\epsilon_{\mathrm{DD}}^2}
		\left(
		2\delta_{jk}
		-
		\delta_{j,k+1}
		-
		\delta_{j,k-1}
		\right)\)
		\\
		\addlinespace[0.3em]
		\(\mathrm{NN}\)
		&
		\(\displaystyle
		m^2\delta_{jk}
		+
		\frac{1}{\epsilon_{\mathrm{NN}}^2}
		\left[
		\bigl(2-\delta_{j1}-\delta_{jN}\bigr)\delta_{jk}
		-
		\delta_{j,k+1}
		-
		\delta_{j,k-1}
		\right]\)
		\\
		\addlinespace[0.3em]
		\(\mathrm{DN}\)
		&
		\(\displaystyle
		m^2\delta_{jk}
		+
		\frac{1}{\epsilon_{\mathrm{DN}}^2}
		\left[
		\bigl(2-\delta_{jN}\bigr)\delta_{jk}
		-
		\delta_{j,k+1}
		-
		\delta_{j,k-1}
		\right]\)
		\\
		\addlinespace[0.3em]
		\bottomrule
	\end{tabular}
\end{table}
Here \(j,k=1,\ldots,N\), with the indices in the PBC case understood cyclically modulo \(N\). All four coupling matrices are real and symmetric. The DD and DN matrices are positive definite for  \(m^2\geq0\), whereas the NN and PBC matrices are positive definite only for \(m^2>0\). At \(m=0\), the latter two contain a zero mode, which we regulate numerically by introducing a small positive mass.

\section{Reality}
\label{app:reality}

For a massless real scalar field on the interval \([0,L]\) with Dirichlet boundary condition, let \(A_0=A_t=A=[0,a]\) denote the same spatial region on the two time slices. 
For \(x,y\in(0,a)\) and \(t>0\), the commutator on the full line is
\begin{equation}
\left[\phi(0,x),\phi(t,y)\right]_{\mathrm{full\,line}}
=
\frac{\mathrm{i}}{2}\,
\Theta\!\left(t-|x-y|\right).
\end{equation}

The Dirichlet boundary conditions are implemented by extending the
field oddly across both endpoints,
\(\phi(t,-x)=-\phi(t,x)\) and
\(\phi(t,2L-x)=-\phi(t,x)\). Thus, the resulting commutator on the interval is
\begin{equation}
\left[\phi(0,x),\phi(t,y)\right]
=
\frac{\mathrm{i}}{2}
\sum_{k\in\mathbb Z}
\left[
\Theta\!\left(t-|x-y+2kL|\right)
-
\Theta\!\left(t-|x+y+2kL|\right)
\right],
\label{eq:DD-image-commutator}
\end{equation}
where the first term in each summand represents direct propagation, while the second represents propagation after reflection from a Dirichlet
boundary. The remaining basic commutators involving
\(\pi=\partial_t\phi\) are obtained by taking the corresponding time
derivatives of this Pauli--Jordan function and therefore vanish whenever
the function vanishes identically throughout the time interval under
consideration.
In the window \(2a<t<2L-2a\), only the \(k=0\) terms in
Eq.~\eqref{eq:DD-image-commutator} contribute, and the corresponding
direct and reflected contributions to the Pauli--Jordan function cancel. Consequently, the Weyl operators associated with these two regions commute, and so do their local operator algebras. It then follows that \(T_{A_0A_t}\) is positive semidefinite, since its
trace pairing with every positive operator on the tensor-product Hilbert
space is nonnegative. Indeed, for an arbitrary operator on the
tensor-product Hilbert space, we have
\begin{equation}
\operatorname{tr}\!\left[
T_{A_0A_t}
\left(\sum_i O_i\otimes O_i'\right)^\dagger
\left(\sum_j O_j\otimes O_j'\right)
\right]
=
\sum_{i,j}
\operatorname{Tr}\!\left[
\rho_0\,
O_i^\dagger O_j\,
O_i^{\prime\dagger}(t)O_j'(t)
\right].
\label{eq:TA0At-positive}
\end{equation}

Since the two local operator algebras commute, the right-hand side of 
Eq.~\eqref{eq:TA0At-positive} can be rearranged as
\begin{equation}
\sum_{i,j}
\operatorname{Tr}\!\left[
\rho_0\,
O_i^\dagger O_i^{\prime\dagger}(t)\,
O_j O_j'(t)
\right]
=
\operatorname{Tr}\!\left[
\rho_0
\left(\sum_i O_iO_i'(t)\right)^\dagger
\left(\sum_j O_jO_j'(t)\right)
\right]
\ge 0 ,
\end{equation}
where the final inequality follows from the positive semidefiniteness
of \(\rho_0\).
The reality and positivity conclusions below also extend to mixed initial states by purification with an auxiliary reference system. For simplicity, we first consider a pure initial state, \(\rho_0=|\psi\rangle\langle\psi|\), and decompose the Hilbert space as \(\mathcal{H}=\mathcal{H}_A\otimes\mathcal{H}_B\), where \(B=\bar A=[a,L]\). Here, Latin and Greek indices refer to the \(A\) and \(B\) factors, respectively. Substituting this expression for \(\rho_0\) into Eq.~\eqref{eq:full-spacetime-density-matrix-general}, we obtain
\begin{equation}
\left(T_{C_0C_t}\right)_{k\rho,l\sigma;\,i\mu,j\nu}
=
\sum_{m,\kappa}
\psi_{m\kappa}\,
\psi_{i\mu}^{*}\,
U_{l\sigma;m\kappa}\,
U_{j\nu;k\rho}^{*}.
\end{equation}

Taking the partial traces over \(B_0B_t\) and \(B_0A_t\), respectively, we obtain
\begin{align}
\left(T_{A_0A_t}\right)_{kl;\,ij}
&=
\sum_{m,\kappa,\lambda,\eta}
\psi_{m\kappa}\,
\psi_{i\lambda}^{*}\,
U_{l\eta;m\kappa}\,
U_{j\eta;k\lambda}^{*},
\\
\left(T_{A_0B_t}\right)_{j\nu;\,i\mu}
&=
\sum_{m,n,\kappa,\lambda}
\psi_{m\kappa}\,
\psi_{i\lambda}^{*}\,
U_{n\nu;m\kappa}\,
U_{n\mu;j\lambda}^{*}.
\end{align}

For later use, we record the matrix elements of the partial transpose with respect to the \(A_t\) factor:
\begin{equation}
\left(T_{A_0A_t}^{T_{A_t}}\right)_{kl;\,ij}
:=
\left(T_{A_0A_t}\right)_{kj;\,il}
=
\sum_{m,\kappa,\lambda,\eta}
\psi_{m\kappa}\,
\psi_{i\lambda}^{*}\,
U_{j\eta;m\kappa}\,
U_{l\eta;k\lambda}^{*}.
\end{equation}

We now use the fact that, for any operators \(F\) and \(G\), the products \(FG\) and \(GF\) have the same nonzero spectrum, even though they may act on different spaces. Here we set
\begin{equation}
F_{k\mu;\,ij}
:=
\delta_{ki}
\sum_{m,\kappa}
\psi_{m\kappa}\,U_{j\mu;m\kappa},
\qquad
G_{jk;\,i\mu}
:=
\sum_{\lambda}
\psi_{i\lambda}^{*}\,
U_{k\mu;j\lambda}^{*},
\end{equation}
which give $T_{A_0B_t}=FG$ and $T_{A_0A_t}^{T_{A_t}}=GF$, and hence 
\begin{equation}
\operatorname{tr}\!\left(T_{A_0B_t}^{\,n}\right)
=
\operatorname{tr}\!\left[
\left(T_{A_0A_t}^{T_{A_t}}\right)^n
\right].
\label{eq:trace-identity}
\end{equation}

Thus, the two
operators appearing in Eq.~\eqref{eq:trace-identity} have the same
nonzero spectrum, although they act on different tensor-product
Hilbert spaces.
Since \(T_{A_0A_t}\) is positive semidefinite, it is Hermitian.
Its partial transpose is therefore also Hermitian and has a real
spectrum. Equation~\eqref{eq:trace-identity} then implies that
\(\operatorname{tr}(T_{A_0B_t}^{\,n})\) is real for every positive
integer \(n\). In particular, all even moments are strictly positive,
since they are sums of even powers of real eigenvalues and cannot
vanish because \(\operatorname{tr}(T_{A_0B_t})=1\).
For a pure Gaussian initial state, the construction in
Sec.~\ref{sec:lattice-framework-timelike-entanglement} shows that 
\(T_{A_0A_t}\) has a Gaussian kernel. Partial
transposition preserves this Gaussian form, so the matrix elements of
its partial transpose have the same form as in
Eq.~\eqref{eq:TAB-Gaussian-matrix-element}. The spectral construction
in Sec.~\ref{subsec:spectrum-and-entropies} therefore applies to
\(T_{A_0A_t}^{T_{A_t}}\), whose nonzero spectrum has the same form as in
Eq.~\eqref{eq:TAB-spectrum}.
The corresponding
parameters \(\widetilde{\xi}_\mu\) are real by Hermiticity and satisfy
\(|\widetilde{\xi}_\mu|<1\). Hence, for every integer \(n\ge2\),
\begin{equation}
\operatorname{tr}\!\left[
\left(T_{A_0A_t}^{T_{A_t}}\right)^n
\right]
=
\prod_{\mu=1}^{d_{AA}}
\frac{(1-\widetilde{\xi}_\mu)^n}{1-\widetilde{\xi}_\mu^{\,n}}
>0.
\end{equation}
Combining this result with Eq.~\eqref{eq:trace-identity} gives
\(\operatorname{tr}(T_{A_0B_t}^{\,n})\) strictly positive for every
integer \(n\ge2\).

\section{Twist-Operator Correlation Functions}

\subsection{Cylinder Four-Point Function}
\label{app:periodic-twist-correlator}

To obtain the Lorentzian twist-operator four-point function for the noncompact massless real scalar field in the vacuum with periodic boundary conditions, we start from the corresponding Euclidean twist-operator four-point function on the plane. Global conformal invariance fixes the latter up to a scaling function. Starting from the complex-scalar result in \cite{Calabrese:2009ez}, we take its decompactification limit and obtain the correlator for a single real scalar by taking the square root. We then map the resulting Euclidean correlator to a cylinder of circumference \(L\) and specialize it to the insertion configuration of interest. Finally, we implement the Schwinger--Keldysh continuation \(\tau_1\to\epsilon_1\) and \(\tau_2\to\mathrm{i}t+\epsilon_2\), with \(\epsilon_1>\epsilon_2>0\). Equivalently, defining \(\tau=\tau_1-\tau_2\), we take \(\tau\to-\mathrm{i}t+(\epsilon_1-\epsilon_2)\) and subsequently send \(\epsilon_1-\epsilon_2\to0^+\). Under this continuation, the two independent Lorentzian cylinder cross ratios are
\begin{equation}
	\begin{aligned}
		x
		&=
		\frac{
			\sin\left(\frac{\pi(b-a)}{L}\right)
			\sin\left(\frac{\pi(d-c)}{L}\right)
		}{
			\sin\left(\frac{\pi(t+a-c+\mathrm{i}0^+)}{L}\right)
			\sin\left(\frac{\pi(t+b-d+\mathrm{i}0^+)}{L}\right)
		},
		\\[0.6em]
		\bar x
		&=
		\frac{
			\sin\left(\frac{\pi(b-a)}{L}\right)
			\sin\left(\frac{\pi(d-c)}{L}\right)
		}{
			\sin\left(\frac{\pi(t+c-a+\mathrm{i}0^+)}{L}\right)
			\sin\left(\frac{\pi(t+d-b+\mathrm{i}0^+)}{L}\right)
		}.
	\end{aligned}
	\label{eq:periodic-lorentzian-cross-ratios}
\end{equation}

The corresponding components of the Lorentzian cylinder kinematic factor are
\begin{equation}
	\begin{aligned}
		R
		&=
		\left(\frac{\pi}{L}\right)^2
		\frac{
			\sin\left(\frac{\pi(t+a-c+\mathrm{i}0^+)}{L}\right)
			\sin\left(\frac{\pi(t+b-d+\mathrm{i}0^+)}{L}\right)
		}{
			\sin\left(\frac{\pi(b-a)}{L}\right)
			\sin\left(\frac{\pi(d-c)}{L}\right)
			\sin\left(\frac{\pi(t+a-d+\mathrm{i}0^+)}{L}\right)
			\sin\left(\frac{\pi(t+b-c+\mathrm{i}0^+)}{L}\right)
		},
		\\[0.8em]
		\bar R
		&=
		\left(\frac{\pi}{L}\right)^2
		\frac{
			\sin\left(\frac{\pi(t+c-a+\mathrm{i}0^+)}{L}\right)
			\sin\left(\frac{\pi(t+d-b+\mathrm{i}0^+)}{L}\right)
		}{
			\sin\left(\frac{\pi(b-a)}{L}\right)
			\sin\left(\frac{\pi(d-c)}{L}\right)
			\sin\left(\frac{\pi(t+d-a+\mathrm{i}0^+)}{L}\right)
			\sin\left(\frac{\pi(t+c-b+\mathrm{i}0^+)}{L}\right)
		}.
	\end{aligned}
	\label{eq:periodic-lorentzian-kinematic-components}
\end{equation}

To express the result compactly, we define \(Q:=\epsilon_{\mathrm{uv}}^4R\bar R\) and \(\Delta_n:=\frac{1}{12}(n-\frac{1}{n})\). Using the result of \cite{Calabrese:2009ez}, we write
\begin{equation}
	\mathcal I_\nu(x,\bar x)
	:=
	\frac{1}{2}
	\left[
	F_\nu(x)F_\nu(1-\bar x)
	+
	F_\nu(\bar x)F_\nu(1-x)
	\right],
\end{equation}
where \(F_\nu(z):={}_2F_1(\nu,1-\nu;1;z)\) and \({}_2F_1(a,b;c;z)\) denotes the Gauss hypergeometric function. In terms of these quantities, the Lorentzian twist-operator four-point function is
\begin{equation}
	\left\langle\sigma_n(a,0)\widetilde{\sigma}_n(b,0)\sigma_n(c,t)\widetilde{\sigma}_n(d,t)\right\rangle_{\mathrm{cyl}}
	=
	\mathcal C_n Q^{\Delta_n}
	\left[
	\prod_{k=1}^{n-1}
	\mathcal I_{k/n}(x,\bar x)
	\right]^{-1/2},
	\label{eq:periodic-lorentzian-four-point-twist}
\end{equation}
where \(\mathcal C_n\) is independent of the insertion points and the time separation. Taking the logarithm then gives the corresponding R\'enyi entropy,
\begin{equation}
	S_n
	=
	c_n
	-
	\frac{n+1}{12n}\log Q
	+
	\frac{1}{2(n-1)}
	\sum_{k=1}^{n-1}
	\log\mathcal I_{k/n}(x,\bar x),
	\label{eq:periodic-lorentzian-renyi}
\end{equation}
where \(c_n\) is a time-independent normalization constant.

\subsection{Strip Two-Point Function}
\label{app:strip-lorentzian-complementary}

To obtain the Lorentzian twist-operator two-point function for the noncompact massless real scalar field in the vacuum on a finite segment with DD or NN boundary conditions, we start from the corresponding Euclidean twist-operator two-point function on the unit disk. Boundary conformal invariance fixes the latter up to a scaling function, whose explicit form is given in \cite{Estienne:2023ekf}. We then map the Euclidean result to a strip of width \(L\) and specialize it to the insertion configuration of interest. Finally, we implement the same Schwinger--Keldysh continuation as in Section~\ref{app:periodic-twist-correlator}. Under this continuation, the Lorentzian strip cross ratio is
\begin{equation}
	r
	=
	-
	\frac{
		\sin^2\left(\frac{\pi(t+\mathrm{i}0^+)}{2L}\right)
	}{
		\sin^2\left(\frac{\pi a}{L}\right)
		-
		\sin^2\left(\frac{\pi(t+\mathrm{i}0^+)}{2L}\right)
	}.
	\label{eq:strip-lorentzian-cross-ratio}
\end{equation}

The arguments of the scaling functions for the two boundary conditions are
\begin{equation}
	x_{\mathrm{DD}}
	=
	1-r,
	\qquad
	x_{\mathrm{NN}}
	=
	r.
	\label{eq:strip-lorentzian-scaling-arguments}
\end{equation}

The corresponding Lorentzian strip kinematic factor is
\begin{equation}
	Q
	=
	-
	\left(
	\frac{2L}{\pi\epsilon_{\mathrm{uv}}}
	\right)^2
	\frac{
		\sin^2\left(\frac{\pi a}{L}\right)
		\sin^2\left(\frac{\pi(t+\mathrm{i}0^+)}{2L}\right)
	}{
		\sin^2\left(\frac{\pi a}{L}\right)
		-
		\sin^2\left(\frac{\pi(t+\mathrm{i}0^+)}{2L}\right)
	}.
	\label{eq:strip-lorentzian-kinematic-factor}
\end{equation}

Using the result of \cite{Estienne:2023ekf}, the Lorentzian twist-operator two-point function can be written as
\begin{equation}
	\left\langle\sigma_n(a,0)\widetilde{\sigma}_n(a,t)\right\rangle_{\mathrm{strip},\alpha}
	=
	\mathcal C_{n,\alpha}Q^{-\Delta_n}
	\left[
	\prod_{k=1}^{n-1}
	F_{k/n}(x_{\alpha})
	\right]^{-1/2},
	\qquad
	\alpha=\mathrm{DD},\mathrm{NN},
	\label{eq:strip-lorentzian-specialized-two-point-twist}
\end{equation}
where \(\Delta_n\) and \(F_\nu\) are defined in Section~\ref{app:periodic-twist-correlator}, and \(\mathcal C_{n,\alpha}\) is independent of the insertion point and the time separation. Taking the logarithm then gives the corresponding R\'enyi entropy,
\begin{equation}
	S_{n,\alpha}
	=
	c_{n,\alpha}
	+
	\frac{n+1}{12n}\log Q
	+
	\frac{1}{2(n-1)}
	\sum_{k=1}^{n-1}
	\log F_{k/n}(x_{\alpha}),
	\qquad
	\alpha=\mathrm{DD},\mathrm{NN},
	\label{eq:strip-lorentzian-renyi}
\end{equation}
where \(c_{n,\alpha}\) is a time-independent normalization constant.

\subsubsection*{Phase Analysis}
\begin{figure}[htbp]
	\centering
	
	\begin{minipage}{0.32\textwidth}
		\centering
		\includegraphics[width=0.95\linewidth]{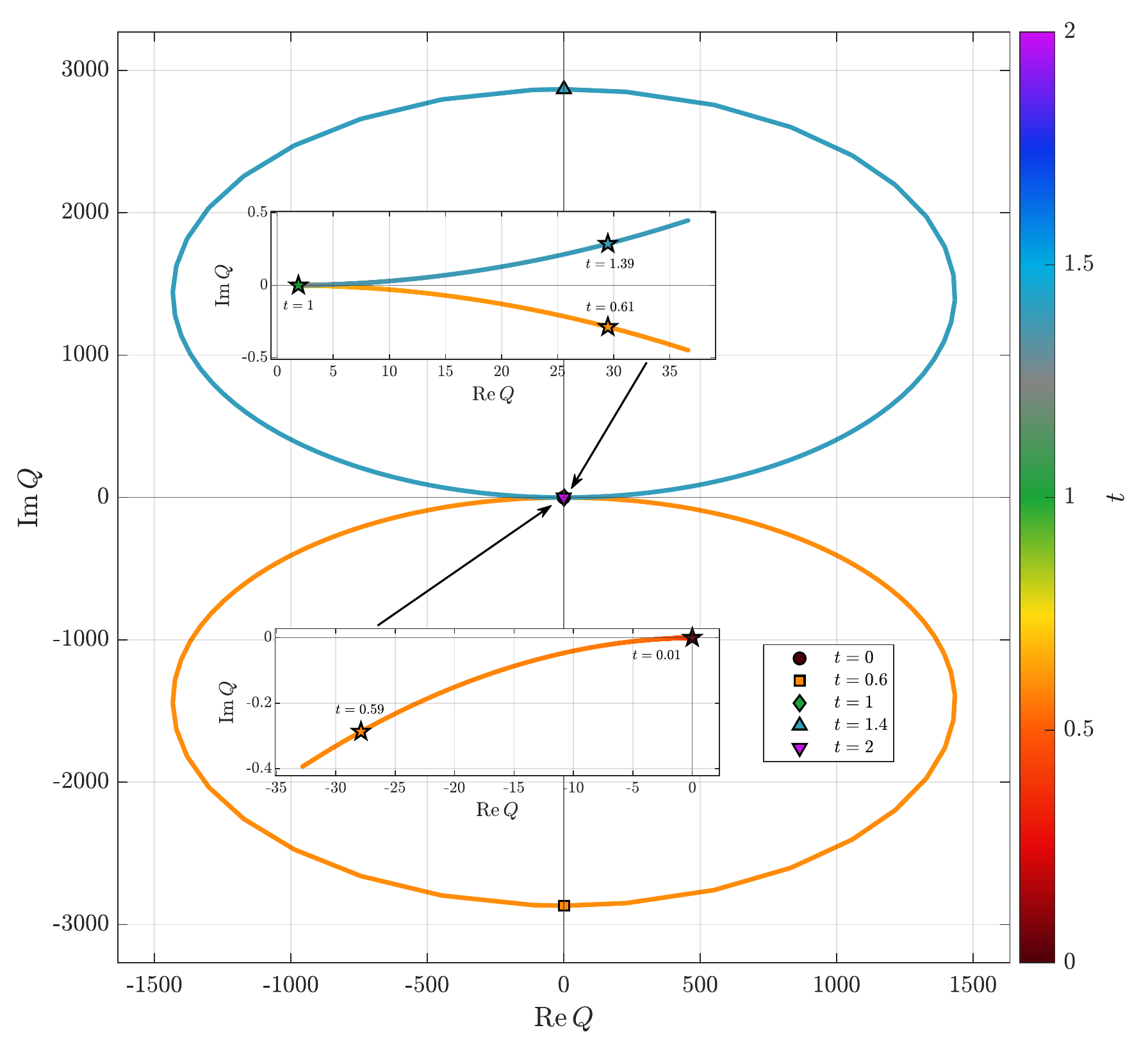}
	\end{minipage}
	\hfill
	\begin{minipage}{0.32\textwidth}
		\centering
		\includegraphics[width=0.95\linewidth]{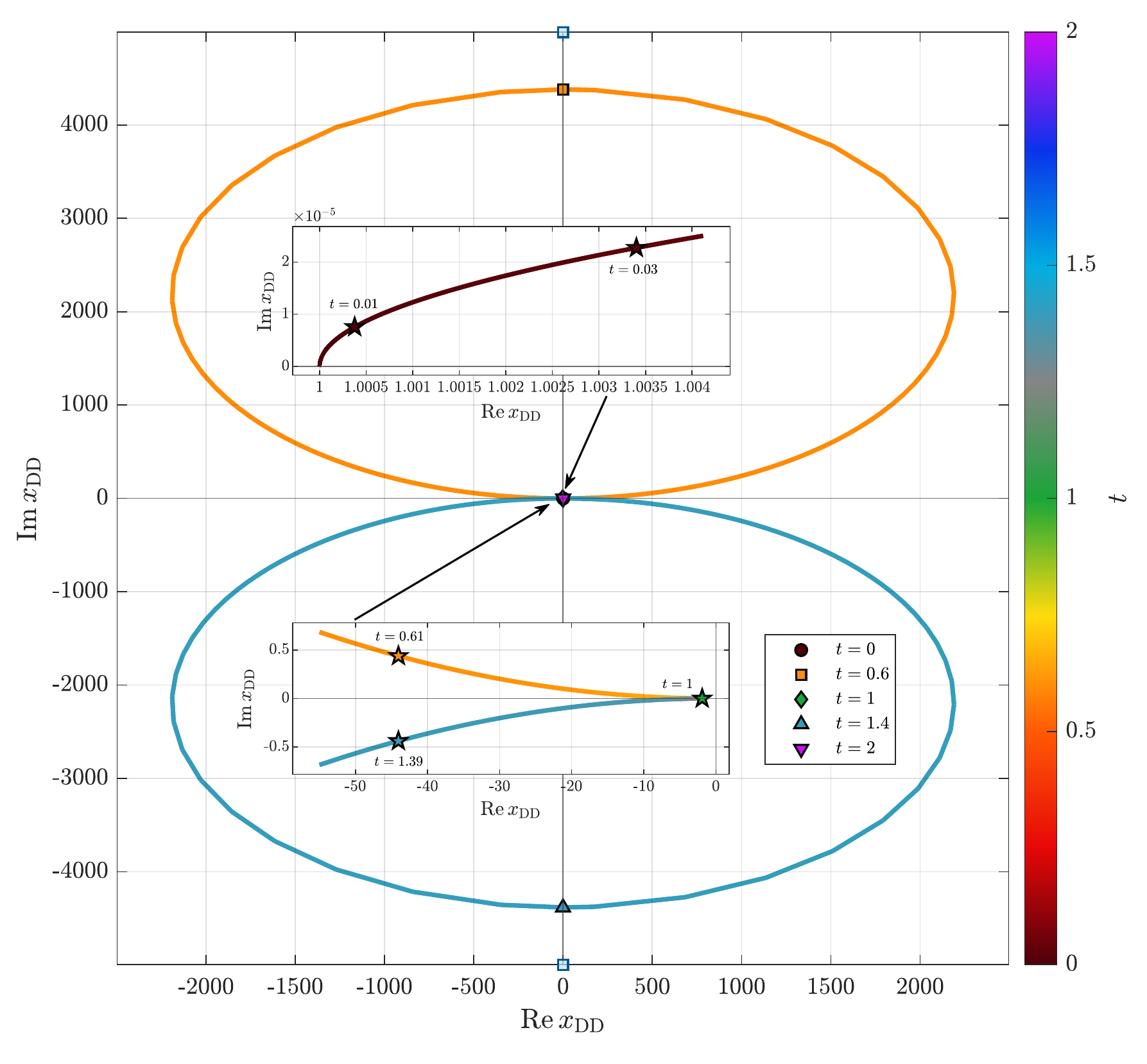}
	\end{minipage}
	\hfill
	\begin{minipage}{0.32\textwidth}
		\centering
		\includegraphics[width=0.95\linewidth]{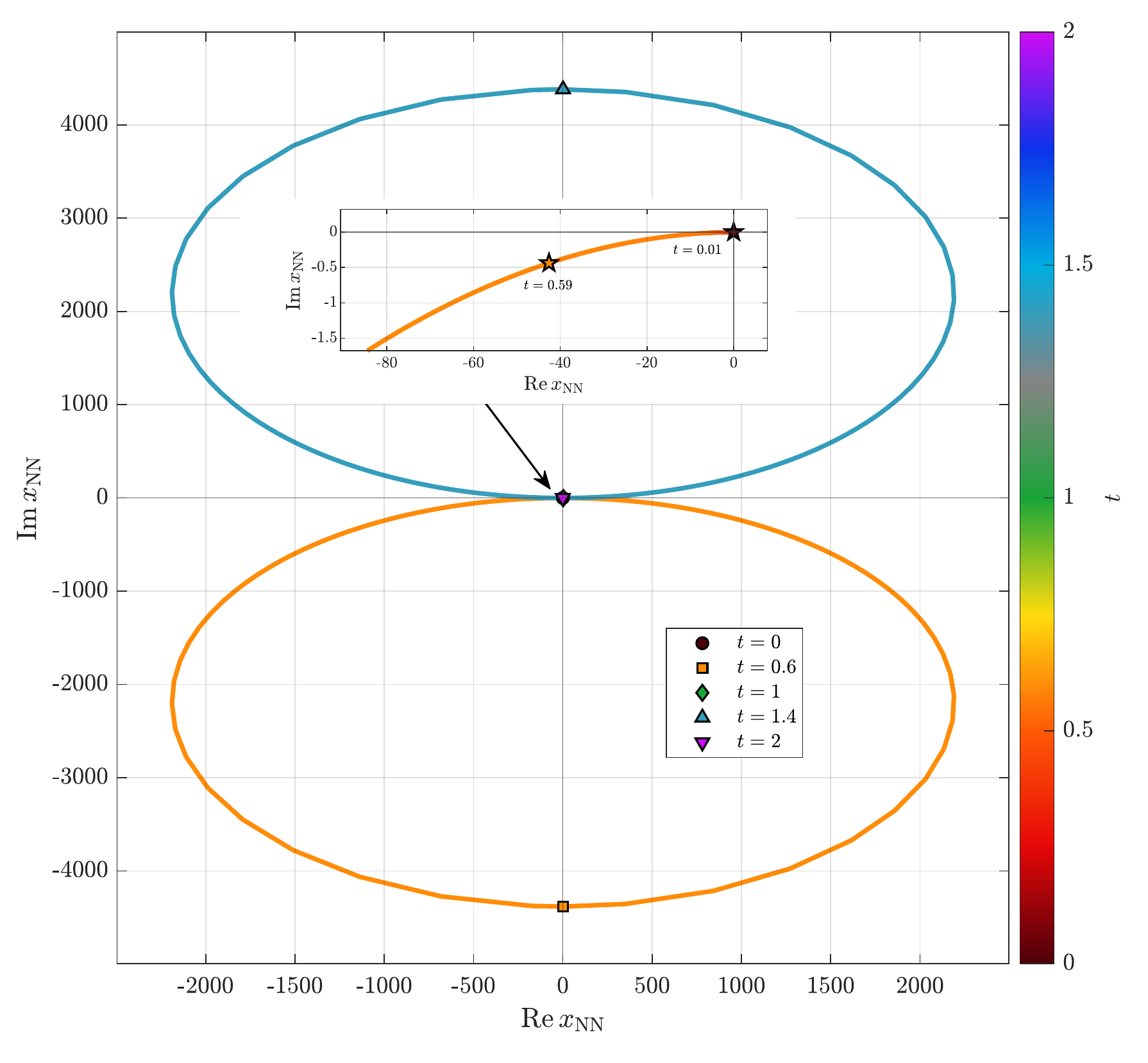}
	\end{minipage}
	
	\par\vspace{1.3em}
	
	\begin{minipage}{0.32\textwidth}
		\centering
		\includegraphics[width=0.90\linewidth]{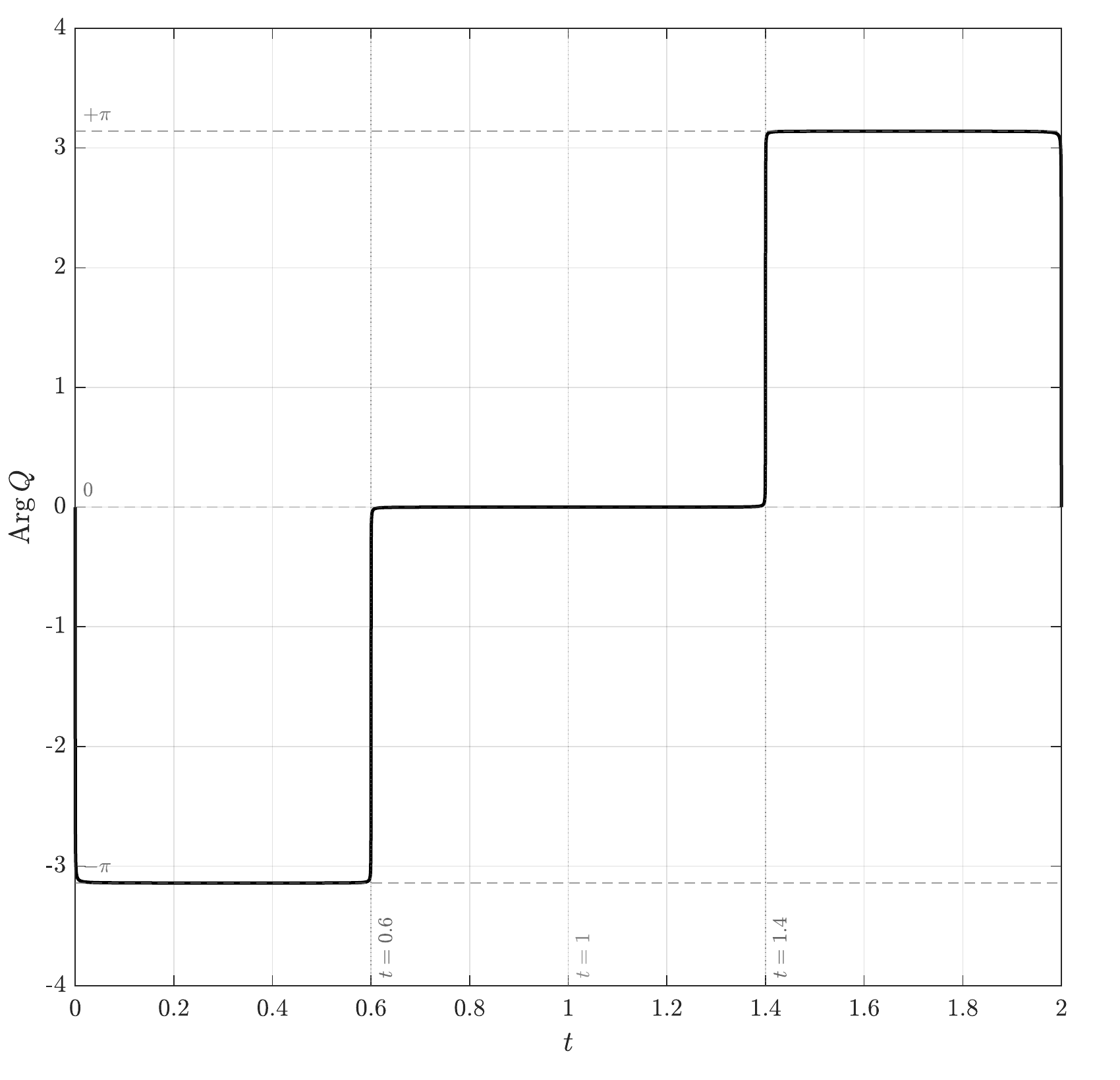}
	\end{minipage}
	\hfill
	\begin{minipage}{0.32\textwidth}
		\centering
		\includegraphics[width=0.90\linewidth]{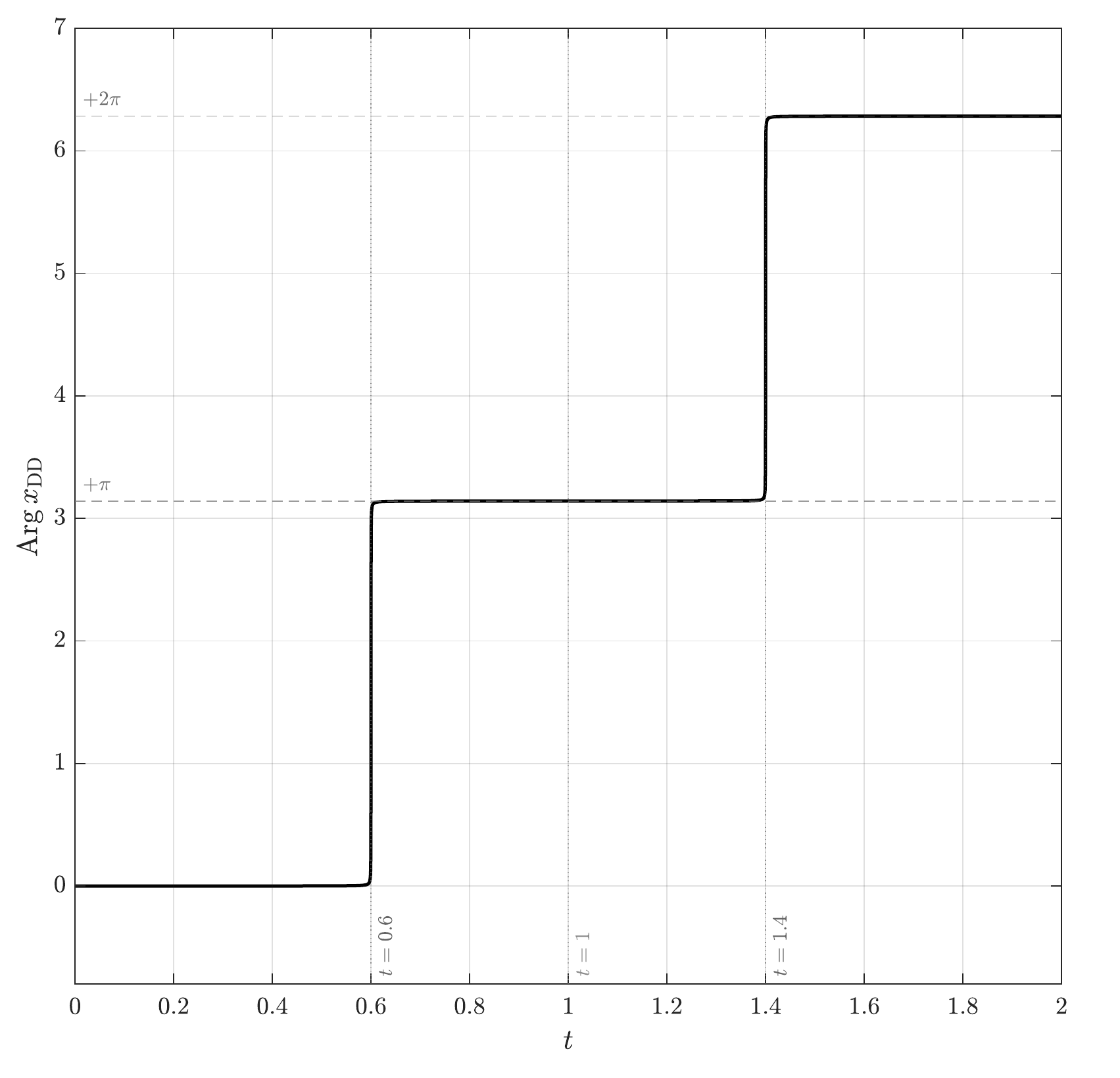}
	\end{minipage}
	\hfill
	\begin{minipage}{0.32\textwidth}
		\centering
		\includegraphics[width=0.90\linewidth]{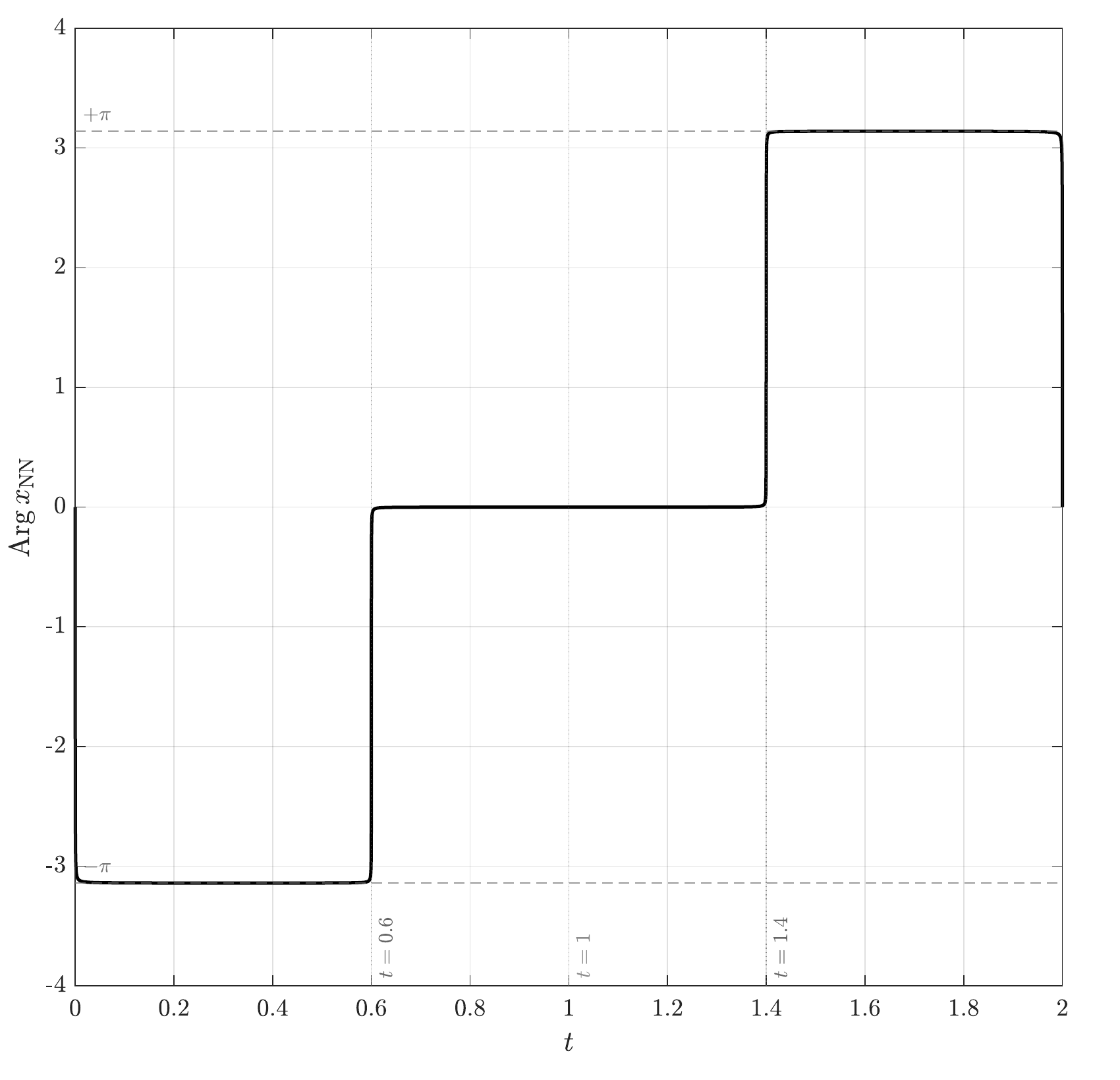}
	\end{minipage}
	
	\caption{Plots of the trajectories and phases of
		\(Q\), \(x_{\mathrm{DD}}\), and \(x_{\mathrm{NN}}\)
		with \(L=1\), \(a=0.3\) and \(\delta=10^{-4}\).}
	\label{fig:trajectories_phases}
\end{figure}

We now explain the vanishing and constant imaginary parts of the BCFT results under the DD and NN boundary conditions, respectively. Although these properties follow directly from \eqref{eq:strip-lorentzian-renyi}, their origin is more transparent from the complex trajectories and continuously tracked phases displayed in Figure~\ref{fig:trajectories_phases}. For this numerical illustration, the \(\mathrm{i}0^+\) prescription implicit in \eqref{eq:strip-lorentzian-renyi} is implemented by a small positive regulator, chosen here as \(\delta=10^{-4}\). All multivalued functions are initialized on their principal branches and subsequently tracked continuously. In the following, we assume \(0<a<\frac{L}{2}\), as in the geometry considered in the main text.

\par\bigskip
\paragraph{DD boundary.}
For \(2a<t<2L-2a\), Fig.~\ref{fig:trajectories_phases} shows that \(Q\) lies close to the positive real axis. Its argument is initialized at \(t=0\) by the principal value and subsequently tracked continuously. Upon removing the regulator, the accumulated argument vanishes throughout this interval; hence, \(\log Q\) is real and does not contribute to the imaginary part of the R\'enyi entropy.

We next consider the hypergeometric function. At \(t=0\), both \(F_{\nu}(x_{\mathrm{DD}})\) and its outer logarithm are initialized on their principal branches. For \(0<t<2a\), the regulated trajectory of \(x_{\mathrm{DD}}\) passes above, but does not cross, the principal branch cut \([1,\infty)\) of the hypergeometric function. It then bends through the upper half-plane and approaches the negative real axis for \(2a<t<L\). Thus, \(F_{\nu}(x_{\mathrm{DD}})\) remains on its principal branch. For \(\nu=k/n\), with \(0<\nu<1\), the Euler integral representation of this branch is
\begin{equation}
	F_{\nu}(x)
	=
	{}_2F_1(\nu,1-\nu;1;x)
	=
	\frac{\sin(\pi\nu)}{\pi}
	\int_0^1
	\mathrm{d}u\,
	u^{-\nu}(1-u)^{\nu-1}(1-xu)^{-\nu}.
	\label{eq:hypergeometric_integral}
\end{equation}

This representation has two relevant consequences. First, for \(x<0\), every factor in the integrand is positive, so \(F_{\nu}(x)>0\). Second, it determines how \(F_{\nu}(x_{\mathrm{DD}})\) approaches the positive real axis. As shown in Fig.~\ref{fig:trajectories_phases}, \(\operatorname{Im}x_{\mathrm{DD}}>0\) for \(2a<t<L\). Therefore, \(1-x_{\mathrm{DD}}u\) lies in the lower half-plane for every \(0<u<1\), whereas its principal power \((1-x_{\mathrm{DD}}u)^{-\nu}\) lies in the upper half-plane. Since all other factors in the integral are positive, \(\operatorname{Im}F_{\nu}(x_{\mathrm{DD}})>0\). For \(L<t<2L-2a\), the signs are reversed: \(\operatorname{Im}x_{\mathrm{DD}}<0\) and \(\operatorname{Im}F_{\nu}(x_{\mathrm{DD}})<0\). At \(t=L\), \(x_{\mathrm{DD}}\) is negative real and \(F_{\nu}(x_{\mathrm{DD}})>0\).

Accordingly, \(F_{\nu}(x_{\mathrm{DD}})\) remains in the upper half-plane until it reaches the positive real axis at \(t=L\), after which it enters the lower half-plane. Its trajectory neither crosses the branch cut of the outer logarithm nor winds around the origin, so the outer logarithm remains on its principal branch. Upon removing the regulator, \(\log F_{k/n}(x_{\mathrm{DD}})\) is real for \(2a<t<2L-2a\). Since the \(\log Q\) term and all hypergeometric terms in \eqref{eq:strip-lorentzian-renyi} are real, \(\operatorname{Im}S_{n,\mathrm{DD}}=0\) throughout this interval.

\par\bigskip
\paragraph{NN boundary.}
For the NN boundary condition, \(Q\) differs from  \(x_{\mathrm{NN}}\) only by a positive real coefficient and therefore has the same phase. Fig.~\ref{fig:trajectories_phases} shows that, for \(0<t<2a\), both trajectories approach the negative real axis from the lower half-plane. Their arguments are initialized at \(t=0\) by their principal values and subsequently tracked continuously. Upon removing the regulator, their accumulated arguments are \(-\pi\) throughout this interval, and hence \(\operatorname{Im}\log Q=-\pi\).

At \(t=0\), both \(F_{\nu}(x_{\mathrm{NN}})\) and its outer logarithm are initialized on their principal branches. For \(0<t<2a\), the trajectory of \(x_{\mathrm{NN}}\) remains below the negative real axis and does not cross the principal branch cut \([1,\infty)\) of the hypergeometric function. Thus, \(F_{\nu}(x_{\mathrm{NN}})\) remains on its principal branch. For \(\nu=k/n\), with \(0<\nu<1\), the integral representation in \eqref{eq:hypergeometric_integral} gives \(F_{\nu}(x)>0\) for \(x<0\). Along the regulated trajectory, \(\operatorname{Im}x_{\mathrm{NN}}<0\), so \(1-x_{\mathrm{NN}}u\) lies in the upper half-plane, whereas its principal power \((1-x_{\mathrm{NN}}u)^{-\nu}\) lies in the lower half-plane. Since all other factors in the integral are positive, \(\operatorname{Im}F_{\nu}(x_{\mathrm{NN}})<0\).

Starting from its positive principal-branch value at \(t=0\), \(F_{\nu}(x_{\mathrm{NN}})\) therefore remains in the lower half-plane and approaches the positive real axis from below. Its trajectory neither crosses the branch cut of the outer logarithm nor winds around the origin, so the outer logarithm remains on its principal branch. Upon removing the regulator, \(\log F_{k/n}(x_{\mathrm{NN}})\) is real for \(0<t<2a\).

Thus, all hypergeometric terms in
\eqref{eq:strip-lorentzian-renyi} are real, whereas the \(\log Q\) term has the constant imaginary part \(-\pi\). Consequently, \(\operatorname{Im}S_{n,\mathrm{NN}}=-(n+1)\pi/(12n)\) throughout \(0<t<2a\).

\subsection{Massive Two-Point Function}
\label{app:massive-two-point-correlator}

To obtain the Lorentzian twist-operator two-point function for the
noncompact massive real scalar field in the vacuum on the infinite line, we start from the corresponding Euclidean twist-operator two-point function on the infinite line. The latter admits a form-factor cumulant expansion derived in \cite{Bianchini:2016mra}. The Lorentzian result is then obtained through the Schwinger--Keldysh continuation \(\ell\to-\mathrm{i}t+0^+\). For a separation \(\ell\), the contribution from the \(2j\)-particle sector is
\begin{equation}
	c_{2j}(\ell,n)
	=
	\frac{2\mathrm{i}n}{j(4\pi)^{2j}}
	\int_{\mathbb R^{2j-1}}
	\mathrm{d}^{\,2j-1}\boldsymbol{x}\,
	K_0\!\left(
	m\ell\,d_{2j}(\boldsymbol{x})
	\right)
	\frac{
		\mathcal F_j\!\left(
		\displaystyle\sum_{s=1}^{j}x_{2s-1},n
		\right)
		\sinh\!\left(
		\displaystyle\sum_{s=1}^{j}x_{2s-1}
		\right)
	}{
		\displaystyle
		\cosh\!\left(
		\frac{1}{2}\sum_{s=1}^{2j-1}x_s
		\right)
		\prod_{s=1}^{2j-1}
		\cosh\!\left(\frac{x_s}{2}\right)
	}.
	\label{eq:massive-cumulant-c2j}
\end{equation}

Here \(m\) is the mass and \(K_0\) denotes the modified Bessel function of the second kind. The function \(d_{2j}(\boldsymbol{x})\) is the positive root determined by
\begin{equation}
	d_{2j}^2(\boldsymbol{x})
	=
	\left[
	1+\sum_{p=1}^{2j-1}\cosh\!\left(\sum_{q=p}^{2j-1}x_q\right)
	\right]^2
	-
	\left[
	\sum_{p=1}^{2j-1}\sinh\!\left(\sum_{q=p}^{2j-1}x_q\right)
	\right]^2.
\end{equation}

The remaining functions are
\begin{equation}
	\mathcal F_j(x,n)
	=
	\sum_{s=1}^{j}
	(-1)^s
	\frac{(2j-1)!}{(j-s)!(j+s-1)!}
	\left[
	f\!\left(2x+(2s-1)\mathrm{i}\pi;n\right)
	-
	f\!\left(2x-(2s-1)\mathrm{i}\pi;n\right)
	\right],
\end{equation}
and
\begin{equation}
	f(x;n)
	=
	-\frac{\sin(\pi/n)}
	{n\left[\cosh(x/n)-\cos(\pi/n)\right]}.
\end{equation}

In terms of these functions, the Lorentzian twist-operator two-point
function is
\begin{equation}
	\log\left\langle\sigma_n(0,0)\widetilde{\sigma}_n(0,t)\right\rangle_{\mathrm{full\,line}}
	=
	\sum_{j=1}^{\infty}c_{2j}(-\mathrm{i}t+0^+,n)
	+	
	2\log\langle\sigma_n\rangle.
	\label{eq:massive-lorentzian-two-point}
\end{equation}

The corresponding R\'enyi entropy is
\begin{equation}
	S_n
	=
	c_n
	+
	\frac{1}{1-n}
	\sum_{j=1}^{\infty}
	c_{2j}(-\mathrm{i}t+0^+,n),
	\label{eq:massive-lorentzian-renyi}
\end{equation}
where \(c_n\) is a time-independent normalization constant.

\section{Numerical Implementation}
\label{app:numerical-implementation}

\subsection{Branch Prescriptions}

The entropy defined in Section~\ref{subsec:review-spacetime-density-matrix} involves a complex logarithm whose branch is not fixed by the definition itself. The same ambiguity therefore arises in Eqs.~\eqref{eq:renyi-factorized-TAB} and
\eqref{eq:von-neumann-entropy-final}. The continuum results in Eqs.~\eqref{eq:periodic-lorentzian-renyi},
\eqref{eq:strip-lorentzian-renyi}, and
\eqref{eq:massive-lorentzian-two-point} likewise contain multivalued
logarithms, while the first two also involve hypergeometric functions with nontrivial branch structure. We therefore specify the branch prescriptions used in our numerical calculations.

\paragraph{Branch tracking.}
For the complex logarithm, we consider two prescriptions. The first uses the principal logarithm independently at each point. The second, which we refer to as branch tracking, is initialized on the principal branch at the first point and subsequently follows the phase continuously as the parameter is varied. Integer multiples of \(2\pi\) are added to or subtracted from the principal phase whenever required by continuity. For example, if a complex number crosses the negative real axis from the lower to the upper half-plane, the principal phase jumps from a value near \(-\pi\) to one near \(+\pi\), whereas branch tracking continues smoothly through \(-\pi\).

For the R\'enyi entropy in Eq.~\eqref{eq:renyi-factorized-TAB} and the logarithms appearing in Eqs.~\eqref{eq:periodic-lorentzian-renyi}, \eqref{eq:strip-lorentzian-renyi}, and
\eqref{eq:massive-lorentzian-two-point}, the two prescriptions give identical results in all cases considered here, and we use branch tracking for these quantities. The von Neumann entropy in
Eq.~\eqref{eq:von-neumann-entropy-final} is treated differently. Since \(\lvert\xi_\mu\rvert<1\), the quantities \(1-\xi_\mu\) do not cross the branch cut of the principal logarithm, whereas tracking
\(\log\xi_\mu\) requires matching and following each root as time varies. For lattices with only a few sites, such tracking is feasible but produces
strong oscillations in both the real and imaginary parts of the entropy and
fails to reproduce the known short-time behavior. As the number of lattice
sites increases, the roots become too densely clustered for reliable
individual tracking. We therefore evaluate all logarithms in Eq.~\eqref{eq:von-neumann-entropy-final} on their principal branches.

\paragraph{Hypergeometric function.}
The continuum results in Eqs.~\eqref{eq:periodic-lorentzian-renyi} and
\eqref{eq:strip-lorentzian-renyi} involve
\(F_\nu(x)={}_2F_1(\nu,1-\nu;1;x)\), whose standard principal branch has a branch cut along \([1,\infty)\) on the real axis. We again compare pointwise evaluation on the principal branch with branch tracking, in which the function is initialized on the principal branch and then followed continuously along the path traced by its argument. The two prescriptions give identical results in all cases considered here except for the periodic case in Eq.~\eqref{eq:periodic-lorentzian-renyi} and the NN case of Eq.~\eqref{eq:strip-lorentzian-renyi}. In these two cases, branch tracking agrees with the lattice calculation, whereas pointwise evaluation on the principal branch does not.

\subsection{Plotting}

Small neighborhoods around the singular times associated with null
separation are omitted from the plots. Since the expressions are
undefined at these times, the curves on the two sides are shown
separately rather than connected across the singularities.

\end{document}